\documentclass[12pt,a4paper]{article}
\usepackage{amsmath,amsfonts,euscript,amssymb}
\usepackage{graphicx}
\DeclareSymbolFont{bfitletters}{OML}{cmm}{bx}{it}
\DeclareSymbolFont{bfitoperators}   {OT1}{cmr} {m}{n}
\DeclareMathSymbol{\bfitomega}{\mathord}{bfitletters}{"21}
\DeclareMathSymbol{\bfitgamma}{\mathord}{bfitletters}{"0D}

\newcommand{\be}{\begin{equation}}
\newcommand{\ee}{\end{equation}}
\newcommand{\bea}{\begin{eqnarray}}
\newcommand{\eea}{\end{eqnarray}}

\begin{document}

\textwidth=150mm
\textheight=200mm
\begin{center}
{\bfseries INTRINSIC TIME IN GEOMETRODYNAMICS: INTRODUCTION AND APPLICATION TO FRIEDMANN COSMOLOGY}
\vskip 15mm
{\bf Alexander Pavlov}
\vskip 5mm
Bogoliubov~Laboratory~for~Theoretical~Physics,
Joint~Institute~of~Nuclear~Research,
~Joliot-Curie~street~6,~Dubna,~141980,~Russia; \\
Institute of Mechanics and Energetics\\
Russian State Agrarian University --\\
Moscow Timiryazev Agricultural Academy, Moscow, 127550, Russia\\
alexpavlov60@mail.ru
\end{center}

\vskip 5mm

\begin{abstract}
An intrinsic local time in Geometrodynamics is obtained with using a scaled Dirac's mapping.
By addition of a background metric, one can construct a scalar field. It is suitable to play a role of
intrinsic time.
Cauchy problem was successfully solved in conformal variables because they are physical ones.
First, the intrinsic time as a logarithm of determinant of spatial metric, was applied to a cosmological
problem by Misner. A global time is exist under condition of constant mean curvature slicing of spacetime.
A volume of hypersurface and the so-called mean York's time are canonical conjugated pair. So, the volume is the intrinsic global time by its sense. The experimentally observed redshift in cosmology is the evidence of its existence. An intrinsic time of homogeneous models is global. The Friedmann equation by its sense ties time intervals. Exact solutions of the Friedmann equation in Standard cosmology with standard and conformal units are presented. Theoretical curves interpolated the Hubble diagram on latest supernovae are expressed in analytical form. The class of functions in which the concordance model is described is Weierstrass meromorphic functions. The Standard cosmological model is used for fitting the modern Hubble diagram. The physical interpretation of the modern data from concept of conformal magnitudes is simpler, so is preferable.

\end{abstract}
\newpage
\tableofcontents
\newpage
\section{Introduction. Physical observables}

The Geometrodynamics is a theory of space and time in its inner essence.
The spatial metric $\gamma_{ij}$ carries information about the inner time.
The intrinsic time for cosmological models is constructed of inner metric characteristic of the space.
The time is to be a scalar relative to diffeomorphisms of changing coordinates of the space.
For this demand we use an idea of bimetric formalism, adding some auxiliary spatial metric.
Thus we naturally come to interpretations of observational data from conformal units concept.
The generalized Dirac's mapping \cite{Dirac} allows to extract the intrinsic time.
The metric $\gamma_{ij}$ is factorized in the inner time factor $\Psi^4$ and the conformal metric
$\tilde\gamma_{ij}$.

The Dirac's mapping reflects the transition to physical (conformal) variables.
In spirit of ideas of these ideas \cite{PP}, the conformal metric is a metric of the space,
where we live and make observations.
The choice of conformal measurement standards allows us to separate the cosmic evolution
of observation devices from the evolution of cosmic objects.
Thus we avoid an unpleasant artefact of expanding Universe and the inevitable problem of Big Bang emerged in
Standard Cosmology. After the procedure of deparametrization is implemented the volume of the Universe occurs
its global intrinsic time. In modern papers \cite{SooYu, IIISooYu, Hoi, Eyo, Vasudev, Ramachandra, Shyam, Huei}
one can see applications of local intrinsic time interval in Geometrodynamics.

Werner Heisenberg in Chapter ``Quantum Mechanics and a Talk with Einstein (1925--1926)'' \cite{Heisenberg}
quoted Albert Einstein's statement:
``{\it But on principle, it is quite wrong to try founding a theory on observable magnitudes alone.
In reality the very opposite happens. It is the theory which decides what we can observe}''.
This conversation between two great scientists about the status of observable magnitudes in the theory
(quantum mechanics or General Relativity) remains actual nowadays.

The supernovae type Ia are used as standard candles to test cosmological models. Recent observations of the supernovae have led cosmologists to conclusion of the Universe filled with dust and mysterious dark energy in frame of Standard cosmology \cite{Riess2004}.
Recent cosmological data on expanding Universe challenge cosmologists in insight of Einstein's gravitation.
To explain a reason of the Universe's acceleration the significant efforts have been applied (see, for example, \cite{Bor, Sz}).

The Conformal cosmological model \cite{PP} allows us to describe the supernova data without Lambda term. The evolution of the lengths in the Standard cosmology is replaced by the evolution of the masses in the Conformal cosmology. It allows to hope for solving chronic problems accumulated in the Standard cosmology. Solutions of the Friedmann differential equation belong to a class of Weierstrass meromorphic functions. Thus, it is natural to use them for comparison predictions of these two approaches. The paper presents a continuation of the article on intrinsic time in Geometrodynamics \cite{Geom}.

\section{ADM variational functional of General\\ Relativity. Notations}

The Einstein's General Relativity was presented in Hamiltonian form a half of century ago \cite{DiracProc}.
Paul Dirac manifested, that four-dimensional symmetry is not fundamental property of the physical world.
Instead of spacetime transformations, one should consider canonical transformations of the phase space
variables.

The ADM formalism, based on the Palatini approach, was developed by Richard Arnowitt, Stanley Deser and Charles Misner in 1959 \cite{ADM}. The formalism supposes the spacetime with interval
$$
{\bf g}=g_{\mu\nu}(t, {\bf x})dx^\mu\otimes dx^\nu
$$
was to foliated into a family of space-like surfaces $\Sigma_t,$ labeled by the {\it time coordinate} $t$,
and with spatial coordinates on each slice $x^i$. The metric tensor of spacetime in ADM form looks like
\be\label{gmunu} (g_{\mu\nu})=
\left(
\begin{array}{cc}
-N^2+N_iN^i& N_i\\
N_j& \gamma_{ij}
\end{array}
\right).
\ee
The physical meaning of the metric components are the following:
the lapse function $N(t; x,y,z)$ defines increment of coordinate time $t$, and the shift vector
$N_i (t; x,y,z)$ defines replacement of coordinates of hypersurface under transition to an infinitesimally
close spacetime hypersurface.

The first quadratic form
\be\label{spacemetric}
{\bfitgamma}=\gamma_{ik}(t, {\bf x})dx^i\otimes dx^k
\ee
defines the induced metric on every slice $\Sigma_t$.
The components of spatial matrix $\gamma_{ij}(t; x,y,z)$ (\ref{spacemetric})
contain three gauge functions
describing the spatial coordinates. The three remaining components describe two polarizations of gravitational
waves and many-fingered time. Thus, we defined the foliation $(\Sigma_t, \gamma_{ij})$.

The group of general coordinate transformations conserving such a foliation was found by Abraham Zel'manov
\cite{Zel}, this group involves the reparametrization subgroup
of coordinate time.
This means that the coordinate time, which is not invariant with respect to gauges, in general case,
is not observable.
A large number of papers were devoted to the choice of reference frames (see, for example, a monograph
\cite{Vlad} and references therein).

The components of the extrinsic curvature tensor $K_{ij}$ of every slice are constructed out of the second
quadratic form of the hypersurface, and can be defined as
\be\label{Kij}
K_{ij}:=-\frac{1}{2}{\pounds}_{\bf n}\gamma_{ij},
\ee
where ${\pounds}_{\bf n}$ denotes the Lie derivative along the $n^\mu$, a time-like unit normal to the slice,
direction.
The components of the extrinsic curvature tensor can be found by the formula
\be\label{Kijdef}
K_{ij}=\frac{1}{2N}\left(\nabla_i N_j+\nabla_j N_i-\dot\gamma_{ij}\right),
\ee
where $\nabla_k$ is a Levi--Civita connection associated with the metric $\gamma_{ij}$
$$\nabla_k \gamma_{ij}=0.$$

The Hamiltonian dynamics of General Relativity is built in an infinite-dimensional degenerated phase space
of 3-metrics $\gamma_{ij} ({\bf x},t)$ and densities of their momenta $\pi^{ij}({\bf x},t)$.
The latter are expressed through the tensor of extrinsic curvature
\be\label{piij}
\pi^{ij} := -\sqrt{\gamma}(K^{ij}-K\gamma^{ij}),
\ee
where we introduced notations:
\be
K^{ij}:=\gamma^{ik}\gamma^{jl}K_{kl},\qquad K:=\gamma^{ij}K_{ij},\qquad
\gamma:=\det||\gamma_{ij}||,\qquad \gamma_{ij}\gamma^{jk}=\delta_i^k.\label{gammadet}
\ee
The Poisson bracket is a bilinear operation on
two arbitrary functionals $F[\gamma_{ij}, \pi^{ij}],$ $G[\gamma_{ij}, \pi^{ij}]$  \cite{T}
\be\label{PB}
\{F, G\}=
\int\limits_{\Sigma_t}\, d^3x\left(\frac{\delta F}{\delta \gamma_{ij}(t,{\bf x})}
\frac{\delta G}{\delta \pi^{ij}(t,{\bf x})}- \frac{\delta G}{\delta \gamma_{ij}(t,{\bf x})}
\frac{\delta F}{\delta \pi^{ij}(t,{\bf x})}\right).
\ee

The canonical variables satisfy to the relation
\be\label{gammapi}
\{\gamma_{ij}(t,{\bf x}),\pi^{kl}(t,{\bf x}')\}=\delta_{ij}^{kl}\delta ({\bf x}-{\bf x}'),
\ee
where
$$
\delta_{ij}^{kl}:=\frac{1}{2}\left(\delta_i^k\delta_j^l+\delta_i^l\delta_j^k\right),
$$
and $\delta ({\bf x}-{\bf x}')$ is the Dirac's $\delta$-function for the volume of $\Sigma_t$.

The super-Hamiltonian of the gravitational field is a functional
\be\label{superHam}
\int\limits_{\Sigma_t}
\left( N{\cal H}_\bot+N^i{\cal H}_i\right)\, d^3x,
\ee
where $N$ and $N^i$ are Lagrange multipliers, ${\cal H}_\bot$, and ${\cal H}_i$ have sense of constraints.
Among them,
\be\label{constraintHam}
{\cal H}_\bot :=
G_{ijkl}\pi^{ij}\pi^{kl}-\sqrt{\gamma}R (\gamma_{ij})
\ee
is obtained from the scalar Gauss relation of the embedding hypersurfaces theory and
called the Hamiltonian constraint. Here $R$ is the Ricci scalar of the space,
$$
 G_{ijkl}:=\frac{1}{2\sqrt{\gamma}}(\gamma_{ik}\gamma_{jl}+\gamma_{il}\gamma_{jk}-\gamma_{ij}\gamma_{kl})
$$
is the supermetric of the 6-dimensional hyperbolic Wheeler -- DeWitt (WDW) superspace \cite{DeWitt}.
Momentum constraints
\be\label{constraintMom}
{\cal H}^i := -2\nabla_j\pi^{ij}
\ee
are obtained from the contracted Codazzi equations of the embedding hypersurfaces theory. They
impose restrictions on possible data $\gamma_{ij}({\bf x},t), \pi^{ij}({\bf x},t)$
on a space-like hypersurface
$\Sigma_t$.
The divergence law, following from (\ref{constraintMom}), is analogous to the Gauss law in Maxwell's
electrodynamics. The Hamiltonian constraint (\ref{constraintHam}) has no analogue in  electrodynamics.
It yields the dynamics of the space geometry itself.
The Hamiltonian dynamics is built of the ADM - variational functional
\be\label{ADMvar}
W=\int\limits_{t_I}^{t_0}
\, dt\int\limits_{\Sigma_t}
d^3x\left(\pi^{ij}\frac{d\gamma_{ij}}{dt}-N{\cal H}_\bot-N^i{\cal H}_i\right),
\ee
where ADM units: $c=1, 16\pi G=1$ were used \cite{ADM}.
The action (\ref{ADMvar}) is obtained of the Hilbert functional after the procedure of $(3 + 1)$ foliation
and the Legendre transformation executed.

These constraints are of the first class, because they identify to the closed algebra
\begin{eqnarray}
&&\{{\cal H}_\bot (t,{\bf x}), {\cal H}_\bot (t,{\bf x}')\}=({\cal H}^i (t,{\bf x})+{\cal H}^i (t,{\bf x}'))
\delta_{,i}({\bf x}-{\bf x}'),\nonumber\\
&&\{{\cal H}_i (t,{\bf x}), {\cal H}_\bot (t,{\bf x}')\}={\cal H}_\bot (t,{\bf x})
\delta_{,i}({\bf x}-{\bf x}'),
\nonumber\\
&&\{{\cal H}_i (t,{\bf x}), {\cal H}_j (t,{\bf x}')\}={\cal H}_i (t,{\bf x}')
\delta_{,j}({\bf x}-{\bf x}')+ {\cal H}_j (t,{\bf x})\delta_{,i}({\bf x}-{\bf x}').\nonumber
\end{eqnarray}
The Poisson brackets between constraints vanish on the constraints hypersurface.
In the presence of matter, described by the energy-momentum tensor $T_{\mu\nu}$, the considered constraints
(\ref{constraintHam}), (\ref{constraintMom}) take the form of Einstein's equations
\be\label{Ham0}
{\cal H}_\bot=\sqrt{\gamma}\left(K_{ij}K^{ij}-K^2\right)-\sqrt{\gamma}R+\sqrt{\gamma} T_{\bot\bot},
\ee
\be\label{mom0}
{\cal H}^i=-2\sqrt{\gamma}\nabla_j\left(K^{ij}-\gamma^{ij}K\right)+\sqrt{\gamma} (T_\bot)^i,
\ee
where
\be\label{Tbotbot}
T_{\bot\bot}:=n^\mu n^\nu T_{\mu\nu}
\ee
is the matter density, and
\be\label{Tboti}
(T_\bot)_i:= n^\mu T_{i\mu}
\ee
is the matter momentum density in a normal observer (Euler observer) reference.
The Hamiltonian constraint (\ref{Ham0}) can be expressed in the momentum variables (\ref{piij}):
\be\label{Hc}
{\cal H}_\bot=\frac{1}{\sqrt\gamma}\left(\pi_{ij}\pi^{ij}-\frac{1}{2}\pi^2\right)-\sqrt\gamma R+
\sqrt\gamma T_{\bot\bot},
\ee
as far as
$$
K^{ij}=-\frac{1}{\sqrt{\gamma}}\left(\pi^{ij}-\frac{1}{2}\pi\gamma^{ij}\right),\quad
\pi_{ij}:=\gamma_{ik}\gamma_{jl}\pi^{kl},\quad
\pi:=\gamma_{ij}\pi^{ij},\quad K=\frac{\pi}{2\sqrt{\gamma}}.
$$
The momentum constraints (\ref{mom0}) in the momentum variables are the following:
\be\label{mc}
{\cal H}^i=-2\nabla_j\pi^{ij}+\sqrt\gamma (T_\bot)^i.
\ee

The Poisson structure (\ref{gammapi}) is degenerated by force of existence of the constraints
(\ref{Hc}), (\ref{mc}).
The reduction of the dynamical system on the level of the constraints, in general case, is an open problem.

\section{Shape dynamics}

A.A. Friedmann in his book \cite{Friedmann}, dedicated to cosmology of the Universe, found the following
remarkable words about the principle of scale invariance:
``{\it ... moving from country to country, we have to change the scale, id est, measured in Russia --
by arshins, Germany -- meters, England -- feet. Imagine that such a change of scale we had to do from point
to point, and then we got the above operation of changing of scale. Scale changing in the geometric world
corresponds, in the physical world, to different ways of measuring of the length... Properties of the world,
are divided into two classes: some are independent of the above said change of scale, better to say, do not
change their shape under any changes of scale, while others under changing of the scale, will change their
shape. Let us agree on their own properties of the world, belonging to the first class, and call scale
invariant. Weyl expands the invariance postulate, adding to it the requirements that all physical laws were
scale-invariant properties of the physical world. Consistent with such an extension of the postulate of
invariance, we have to demand that the world equations would be expressed in a form satisfactory to not only
coordinate, but the scale invariance}''.
Radiative breaking of conformal symmetry in a conformal-invariant version of the Standard Model of elementary particles is considered in \cite{Euro}. The fruitful idea of initial conformal symmetry of the theory lead to
right value of Higgs boson mass.

The Einstein's theory of general relativity is covariant under general coordinate transformations.
The group of transformations is an infinite-parameter one. The action of the group can be reduced to
alternating actions of its two finite-parameter subgroups: the spatial linear group $SL (3,1)$ and the
conformal group $SO(4,2)$.
According to Ogievetsky's theorem \cite{Og}, the invariance under the infinite-parameter generally covariant
group is equivalent to simultaneous invariance under the affine and the conformal group.
Using an analogy with phenomenological chiral Lagrangians \cite{VP}, it is possible to obtain phenomenological
affine Lagrangian as nonlinear joint realization of affine and conformal symmetry groups. A nonlinear
realization  of the affine group leads to a symmetry tensor field as a Goldstone field. The requirement that
the theory correspond simultaneously to a realization of the conformal group as well leads uniquely to the
theory of a tensor field whose equations are Einstein's  \cite{BO}. The gravitational field by its origin is a Goldstone field. York's method of decoupling of the
momentum and Hamiltonian constraints \cite{York} is derived on a basis of a mathematical discovery.
It is not an occasion: The physical principle of York's method was based on conformal and affine symmetry that
initially were contained in the theory as artefact.

For recovering an initial conformal symmetry to the space one uses an artificial method.
One can possible to change the gauge symmetry of General Relativity (spatial diffeomorphisms and local
changing of slicing) by gauge symmetry of dynamics of form (spatial diffeomorphisms and local scaling conserved
the global slicing) of spacetime \cite{Foster}.
Following to \cite{Foster}, let us define a class of metrics $\bfitgamma$ of some hypersurface $\Sigma$,
that conserve its volume
$$
{V_\gamma}=\int_\Sigma\, d^3x\sqrt{\gamma (x)},
$$
with help of conformal mapping of metric coefficients
\be
\label{conformalmetric}
\gamma_{ij}(x)\to\exp (4\hat{\phi}(x))\gamma_{ij}(x).
\ee
Here the function is defined
\begin{equation}\label{hatphi}
\hat{\phi}({x}):=\phi(x)-\frac{1}{6}\ln <e^{6\phi}>_{\gamma},
\end{equation}
and operation of meaning by a hypersurface $\Sigma$ for some scalar field $f$
\begin{equation}
<f>_\gamma:=\frac{1}{V_\gamma}\int_\Sigma\,d^3x\sqrt{\gamma (x)}f(x).
\end{equation}
There was introduced St$\ddot{\rm u}$ckelberg scalar field \cite{Stuk} in space by analogy with Deser's
\cite{Deser} introducing of dilaton Dirac field \cite{Diracdilaton} in spacetime and an averaging of
functions \cite{BPZZ, ABNPBPZ} for an arbitrary manifold.

The theorem: {\it conformal mapping (\ref{conformalmetric}) conserves a volume of every hypersurface} was
proved in \cite{Foster}. From the definition (\ref{hatphi}) the conformal factor is expressed
\be
e^{4\hat\phi}=\frac{e^{4\phi}}{\left[<e^{6\phi}>_\gamma\right]^{2/3}}.
\ee
Then Jacobian of transformation is transformed by the formula
$$
\sqrt\gamma\to\frac{e^{6\phi}}{<e^{6\phi}>_\gamma}\sqrt\gamma.
$$
The variation of Jacobian and, correspondingly, of the volume of hypersurface are
$$
\delta\sqrt\gamma=\frac{1}{2}\sqrt\gamma\gamma^{ab}\delta\gamma_{ab},\quad
{\delta V_\gamma}=\frac{1}{2}\sqrt{\gamma (y)}\gamma^{ij}(y){\delta\gamma_{ij}(y)}.
$$
The volume of the hypersurface $V_\gamma$ is conserved
\be
V_\gamma=\int_{\Sigma_t}d^3x\,\sqrt{\gamma}(x)\to\frac{1}{<e^{6\phi}>_\gamma}
\int_{\Sigma_t}d^3x\,e^{6\phi}\sqrt{\gamma}(x)
=V_\gamma\frac{<e^{6\phi}>_\gamma}{<e^{6\phi}>_\gamma}=V_\gamma.\nonumber
\ee
Q.E.D.

The phase space (cotangent bundle over $Riem (\Sigma)$) can be extended with the scalar field $\phi$ and
canonically conjugated momentum density $\pi_\phi$.
Let ${\cal C}$ is a group of conformal transformations of the hypersurface $\Sigma$:
$$(\gamma_{ij},\pi^{ij}; \phi,\pi_\phi)\mapsto (\Gamma_{ij},\Pi^{ij}; \Phi,\Pi_\phi),$$
parameterized by scalar field $\phi$ with a generation functional \cite{Gomes}
\be\label{genF}
F_\phi [\gamma_{ij}, \Pi^{ij}, \phi, \Pi_\phi]:=
\int_\Sigma\, d^3x\left[\gamma_{ij}({x})e^{4\hat{\phi}({x})}\Pi^{ij}({x})+\phi({x})\Pi_\phi({x})\right].
\ee

The canonical transformations in the extended phase space
$$(\gamma_{ij},\pi^{ij};\phi,\pi_\phi)\in\Gamma_{Ext}:=\Gamma\times T^*({\cal C})$$
with the canonical Poisson bracket are the following:
\bea
&&\gamma_{ij}\to{\cal T}_\phi\gamma_{ij}(x):=\Gamma_{ij}=e^{4\hat\phi (x)}\gamma_{ij},\label{can1}\\
&&\pi^{ij}\to{\cal T}_\phi\pi^{ij}(x):=\Pi^{ij}=\nonumber\\
&&e^{-4\hat\phi (x)}\left(
\pi^{ij}(x)-\frac{1}{3}\gamma^{ij}\sqrt\gamma <\pi>(1-e^{6\hat\phi})\right),\label{can2}\\
&&\phi (x)\to{\cal T}_\phi \phi (x):=\Phi=\phi (x),\label{can3}\\
&&\pi_\phi(x)\to{\cal T}_\phi\pi_\phi (x):=\Pi_\phi=
\pi-4\left(\pi_\phi (x)-<\pi>\sqrt\gamma\right).\label{can4}
\eea

After the canonical transformations were implemented, the constraints (\ref{Ham0}), (\ref{mom0})
got the following form:
\bea
\left({\cal H}_\bot\right)_\phi&=&
\frac{e^{-6\hat\phi}}{\sqrt\gamma}
\left(\pi^{ij}\pi_{ij}+\frac{1}{3}\sqrt\gamma\left(1-e^{6\hat\phi}\right)
<\pi>\pi-\right.\nonumber\\
&-&\left.
\frac{1}{6}\gamma\left(1-e^{6\hat\phi}\right)^2<\pi>^2-\frac{1}{2}\pi^2\right)-
\sqrt\gamma\left(Re^{2\hat\phi}-8e^{\hat\phi}\Delta e^{\hat\phi}\right),\nonumber\\
\left({\cal H}^i\right)_\phi&=&
-2e^{-4\hat\phi}\left(\nabla_j\pi^{ij}-
2(\pi-\sqrt\gamma <\pi>)\nabla^i\phi\right),\nonumber\\
{\cal Q}_\phi&=&\pi_\phi-4(\pi-<\pi>\sqrt\gamma).\nonumber
\eea

\section{Cauchy problem in Conformal gravitation}

Let us proceed the solution of the Cauchy problem following to York in conformal variables (denoted by bar)
in detail. ``{\it Note that the configuration space that one is led to by the initial-value equations is not
superspace (the space of Riemannian three-geometries), but ``conformal superspace'' [the space of which each
point is a conformal equivalence class of Riemannian three-geometries]$\times$[the real line]
(i.e., the time $T$)}'' \cite{York}.

The matter characteristics under conformal transformation
\begin{equation}\label{Psifactor}
\gamma_{ij}:=e^{-4\hat\phi}\bar\gamma_{ij}\equiv\Psi^4\bar\gamma_{ij}
\end{equation}
are transformed according to their conformal weights. We denote the transformed matter characteristics
(\ref{Tbotbot}) and (\ref{Tboti}) as
\begin{equation}\label{tildeT}
\bar{T}_{\bot\bot}:=\Psi^8 T_{\bot\bot};\qquad
(\bar{T}_\bot)^i:=\Psi^{10}({T}_\bot)^i.
\end{equation}

After the traceless decomposition of $K^{ij}$
\be\label{KijAijK}
K^{ij}=A^{ij}+\frac{1}{3}K\gamma^{ij},\qquad \gamma_{ij}A^{ij}=0,
\ee
we decompose the traceless part of $A^{ij}$ according to
\be\label{Aijhat}
A^{ij}=\Psi^{-10}\bar{A}^{ij}.
\ee
Then, we obtain conformal variables
$$
\bar{A}^{ij}:=\Psi^{10}A^{ij},\qquad \bar{A}_{ij}:=\bar\gamma_{ik}\bar\gamma_{jl}\bar{A}^{kl}=
\Psi^{-8}\gamma_{ik}\gamma_{jl}\bar{A}^{kl}=\Psi^2 A_{ij}.
$$
The Hamiltonian constraint (\ref{Ham0}) in the new variables
\be
\bar\Delta\Psi-\frac{1}{8}\bar{R}\Psi+\frac{1}{8}\bar{A}_{ij}\bar{A}^{ij}\Psi^{-7}
-\frac{1}{12}K^2\Psi^5+
\frac{1}{8}\bar{T}_{\bot\bot}\Psi^5=0
\label{H}
\ee
is named the Lichnerowicz -- York equation \cite{LYork}.
Here $\bar\Delta:=\bar\nabla_i\bar\nabla^i$ is the conformal Laplacian,
$\bar\nabla_k$ is the conformal connection associated with the conformal metric $\bar\gamma_{ij}$
$$\bar\nabla_k\bar\gamma_{ij}=0,$$
$\bar{R}$ is the conformal Ricci scalar expressed of the Ricci scalar $R$:
\be\label{Rscalar}
R= \Psi^{-4}\bar{R}-8 \Psi^{-5}\bar\Delta\Psi.
\ee
Lichnerowicz originally considered the differential equation (\ref{H})
without matter and in case of a maximal slicing gauge $K=0$ \cite{Lich}.

The momentum constraints (\ref{mom0}) after the decomposition (\ref{Aijhat}) take the form:
\be\label{M}
\bar\nabla_j\bar{A}^{ij}-\frac{2}{3}\Psi^6\bar\nabla^i K+\frac{1}{2}(\bar{T}_\bot)^i=0.
\ee

To solve the Cauchy problem in the General Relativity, York elaborated the conformal
transverse -- traceless method
\cite{YorkSources}.
He made the following decomposition of the traceless part $\bar{A}_{ij}$:
\be\label{longtrans}
\bar{A}^{ij}=\left(\bar{\mathbb L}{X}\right)^{ij}+\bar{A}_{TT}^{ij},
\ee
where $\bar{A}_{TT}^{ij}$ is both traceless and transverse with respect to the metric $\bar\gamma_{ij}$:
$$\bar\gamma_{ij}\bar{A}_{TT}^{ij}=0,\qquad \bar\nabla_j\bar{A}_{TT}^{ij}=0,$$
$\bar{\mathbb L}$ is the {\it conformal Killing operator}, acting on the vector field ${\bf X}$:
\be\label{Killing}
\left(\bar{\mathbb L} {X}\right)^{ij}:=\bar\nabla^i X^j+\bar\nabla^j X^i-
\frac{2}{3}\bar\gamma^{ij}\bar\nabla_k X^k.
\ee
The symmetric tensor $\left(\bar{\mathbb L} {X}\right)^{ij}$ is called the {\it longitudinal part} of
$\bar{A}^{ij}$, whereas
$\bar{A}_{TT}^{ij}$ is called the {\it transverse part} of $\bar{A}^{ij}$.

Using the York's longitudinal-transverse decomposition (\ref{longtrans}), the constraint equation (\ref{H})
can be rewritten in the following form
\be
\bar\Delta\Psi-\frac{1}{8}\bar{R}\Psi+
\frac{1}{8}\left[\left(\bar{\mathbb L} {X}\right)_{ij}+
\bar{A}_{ij}^{TT}\right]\left[\left(\bar{\mathbb L} {X}\right)^{ij}+
\bar{A}_{TT}^{ij}\right]\Psi^{-7}-
\frac{1}{12}K^2\Psi^5+
\frac{1}{8}\bar{T}_{\bot\bot}\Psi^5=0,\label{HYork}
\ee
where the following notations are utilized
$$\left(\bar{\mathbb L} {X}\right)_{ij}:=
\bar\gamma_{ik}\bar\gamma_{jl}\left(\bar{\mathbb L} {X}\right)^{kl},\qquad
\bar{A}_{ij}^{TT}:=\bar\gamma_{ik}\bar\gamma_{jl}\bar{A}_{TT}^{kl};$$
and the momentum equations (\ref{M}) are:
\be\label{MYork}
\bar\Delta_{\mathbb L}X^i-\frac{2}{3}\Psi^6\bar\nabla^i K + \frac{1}{2}(\bar{T}_\bot)^i=0.
\ee
The second order operator $\bar\nabla_j\left(\bar{\mathbb L} {X}\right)^{ij}$, acting on the vector
${\bf X}$, is the
{\it conformal vector Laplacian} $\bar\Delta_{\mathbb L}$:
\be
\bar\Delta_{\mathbb L} X^i:=\bar\nabla_j\left(\bar{\mathbb L} X\right)^{ij}=
\bar\nabla_j\bar\nabla^j X^i+\frac{1}{3}\bar\nabla^i\bar\nabla_j X^j+\bar{R}^i_j X^j.\label{Laplacian}
\ee
To obtain the formula (\ref{Laplacian}), we have used the contracted Ricci identity.

The part of the initial data on $\Sigma_0$ can be freely chosen and other part is constrained, {\it id est}
determined from the
constrained equations (\ref{HYork}), (\ref{MYork}).
One can offer a constant mean curvature condition on Cauchy hypersurface $\Sigma_0$:
\be\label{piT}
K\equiv \frac{\pi}{2\sqrt{\gamma}}=\rm{const}.
\ee
Then the momentum constraints (\ref{mom0}) are separated of the Hamiltonian constraint (\ref{Ham0}) and reduce
to
\be\label{Poisson}
\bar\Delta_{\mathbb L}X^i+\frac{1}{2} (\bar{T}_\bot)^i=0.
\ee
Therefore, we obtain the {\it conformal vector Poisson equation}.
It is solvable for closed manifolds, as it was proved in \cite{ChBIYork}. So, we have

$\bullet$
Free initial data:

conformal factor $\Psi^4$;
conformal metric $\bar\gamma_{ij}$;
transverse tensor $\bar{A}_{TT}^{ij}$;
conformal matter variables $(\bar{T}_{\bot\bot}, (\bar{T}_\bot)^i)$.

$\bullet$
Constrained data:

scalar field $K$;
vector ${\bf X}$, obeying the linear elliptic equations (\ref{MYork}).

Note, after solving the Cauchy problem, we are not going to return to the initial variables,
in opposite to the York's approach. Cauchy problem was successfully solved not by chance after mathematically
formal transition to conformal variables. The point is that we have found just the physical variables.

According to Yamabe's theorem \cite{Yamabe}, {\it any metric of compact Riemannian manifold of dimension more
or equal three, can be transformed to a metric of space with constant scalar curvature.}
If we multiply the equation (\ref{Rscalar})
to $\Psi^{-1}$ and integrate it over all manifold $\Sigma$ we get:
\be
\int_\Sigma\,d^3x\sqrt\gamma R\Psi^{-1}=\int_\Sigma\, d^3x\sqrt{\bar\gamma}(\bar{R}\Psi-8\bar\Delta\Psi )
\equiv
\int_\Sigma\,d^3x\sqrt{\bar\gamma}\bar{R}\Psi.\nonumber
\ee
For compact manifolds, the integral of Laplacian of the scalar function is equal to zero. Consequently,
a sign of scalar of curvature is conserved under conformal transformations. There is a Yamabe's constant
\be
{\textsf{y}}[\Sigma,\gamma ]=\inf_\Psi\left\{\frac{\int d^3x\sqrt\gamma (\Psi^2 R-8\Psi\Delta\Psi )}
{\int d^3x\sqrt\gamma \Psi^6}\right\}.
\ee
According to Yamabe's theorem, in conformal equivalent class of metrics there are metrics,
where a minimum is realized, and hence, they correspond to spaces of constant scalar curvature.
Riemannian metrics are classified according to a sign of Yamabe constant. They are classified according to their positive, negative and zeroth signs.

Multiplying the Lichnerowicz -- York equation (\ref{H}) to $8\Psi^7$, one can rewrite it in the following form
\be\label{LYm}
8\Psi^7\bar\Delta\Psi=\left(\frac{2}{3}K^2-\bar{T}_{\bot\bot}\right)\Psi^{12}+\bar{R}\Psi^8-
\bar{A}_{ij}\bar{A}^{ij}.
\ee
Then we denote
$z\equiv\Psi^4$
and present the right side of the equation (\ref{LYm}) as a polynomial
\be\label{polynomial}
f(z)=\left(\frac{2}{3}K^2-\bar{T}_{\bot\bot}\right) z^3+\bar{R} z^2-\bar{A}_{ij}\bar{A}^{ij}.
\ee
According to the theory of quasilinear elliptic equations, the equation
has a unique solution, if the polynomial of the third order by $z$ (\ref{polynomial})
$$f(z)=\left(\frac{2}{3}K^2-\bar{T}_{\bot\bot}\right)(z-z_1)(z-z_2)(z-z_3)$$
has a unique real root $z_i$.
A generic solution of the York's problem of initial problem is valid in spacetime near the
Cauchy hypersurface of constant mean curvature $(CMC).$
We restrict our consideration here by matter sources adopted to the theorem of existence.

\section{Many-fingered intrinsic time in\\ Geometrodynamics}

Dirac, in searching of dynamical degrees of freedom of the gravitational field, introduced {\it conformal
field variables} $\tilde\gamma_{ij}, \tilde\pi^{ij}$ \cite{Dirac}
\begin{equation}\label{hatgamma}
\tilde\gamma_{ij}:=\frac{\gamma_{ij}}{\sqrt[3]{\gamma}}, \qquad
\tilde\pi^{ij}:=\sqrt[3]{\gamma}\left(\pi^{ij}-\frac{1}{3}\pi\gamma^{ij}\right),
\end{equation}
id est, our choice of the conformal factor (\ref{Psifactor}):
\be\label{fixshape}
\Psi=\gamma^{1/12}.
\ee
Among them, there are only five independent pairs $(\tilde\gamma_{ij}, \tilde\pi^{ij})$ per space point,
because
of unity of the determinant of conformal metric and traceless of the matrix of conformal momentum densities
$$\tilde\gamma:=\det ||\tilde\gamma_{ij}||=1,\qquad \tilde\pi:=\tilde\gamma_{ij}\tilde\pi^{ij}=0.$$
The remaining sixth pair $(D, \pi_D)$
\begin{equation}\label{hatTpT}
D:=-\frac{1}{3}\ln\gamma,\qquad \pi_D:=\pi
\end{equation}
is canonically conjugated. The essence of the transformation (\ref{hatgamma}) lies in the fact, that
the metric $\tilde\gamma_{ij}$ is equal to the whole class of conformally equivalent Riemannian
three-metrics $\gamma_{ij}$. So, the conformal variables describe dynamics of shape of the hypersurface of
constant volume.
The conformal mapping is not a coordinate diffeomorphism. Under the conformal transformation any
angles between vectors are equal and ratios of their
magnitudes are preserved in points with equal coordinates of the spaces \cite{Eisenhart}.
The extracted canonical pair $(D, \pi_D)$ (\ref{hatTpT}) has a transparent physical sense, {viz}:
an intrinsic time $D$ and a Hamiltonian density of gravitation field $\pi_D$.

The concepts of
{\it Quantum Geometrodynamics} of Bryce DeWitt \cite{DeWitt} and John Wheeler \cite{Wheeler}
follow from the Dirac's transformations (\ref{hatgamma})--(\ref{hatTpT}).
They thought the cosmological time to be identical the cosmological scale factor.
Information of time must be contained
in the internal geometry, and Hamiltonian must be given by the characteristics of external geometry (\ref{Kij}).
The canonical variable $D$ (\ref{hatTpT}) plays a role of intrinsic time in Geometrodynamics.

However, generally, as we see from (\ref{hatgamma})-(\ref{hatTpT}), $D$ is not a scalar,
and $\tilde\gamma_{ij}$ is not a tensor under group of diffeomorphisms.
According to terminology in \cite{Zel}, $D$ is not a kinemetric invariant,
$\tilde\gamma_{ij}$ is not a kinemetric invariant 3-tensor under kinemetric group of transformation.
The Dirac's transformations (\ref{hatgamma})--(\ref{hatTpT}) have a limited range of applicability:
they can be used in the coordinates with dimensionless metric determinant.

To construct physical quantities, an approach based on Cartan differential forms, invariant under
diffeomorphisms has been elaborated \cite{ABNPBPZ}.
In the present paper, to overcome these difficulties we use a fruitful idea of {\it bimetric formalism}
\cite{Rosen}.
Spacetime bi-metric theories were founded on some auxiliary background non - dynamical metric
with coordinate components
$f_{ij}({\bf x})$ of some 3-space, Lie-dragged along the coordinate time evolution vector
\begin{equation}\label{dfij}
\frac{\partial f_{ij}}{\partial t}=0.
\end{equation}
A background flat space metric was used for description of asymptotically flat spaces.
It was connected with Cartesian coordinates which are natural for such problems.
The development of these theories has been initiated by the problem of energy in Einstein's
theory of gravitation. The restriction (\ref{dfij}) to the background metric is not strong from mathematical
possibilities, so there is a possibility of its choosing from physical point of view.
If a topology of the space is $S^3$ so the background metric is a metric of sphere; for $S^2\times S^1$ --
the metric corresponding to Hopf one; and for $S^1\times S^1\times S^1$ the metric of compactified flat space should be taken.
The space bi-metric approach is natural for solution of cosmological
problems also, when a static tangent space will be taken in capacity of a background space.

To use an auxiliary metric for a generic case of spatial manifold with arbitrary topology,
let us take a local tangent space  ${\cal T}(\Sigma_t)_{\bf x}$ as a background space
for every local region of our manifold $(\Sigma_t)$.
In every local tangent space we define a set of three linear independent vectors $e^i_{a}$ (dreibein),
numerated by the first Latin indices $a, b$.
The components of the background metric tensor in the tangent space:
$$e^i_{a}e_{b i}=f_{a b}.$$
Along with the dreibein, we introduce three mutual vectors $e^{a}_i$ defined by the orthogonal conditions
$$e_i^{a}e^i_{b}=\delta_b^a,\qquad e_i^{a}e^j_{a}=\delta_i^j.$$
Then, we can construct three linear independent under diffeomorphisms Cartan forms \cite{VP}
\begin{equation}\label{dif}
\omega^{a}(d)=e^{a}_i dx^i.
\end{equation}
The background metric is defined by the differential forms (\ref{dif}):
\begin{equation}
{\bf f}=f_{a b}\omega^{a}(d)\otimes\omega^{b}(d)=f_{ij}dx^i\otimes dx^j,
\end{equation}
where the components of the background metric tensor in the coordinate basis are
$$f_{ij}=f_{a b}e^{a}_i e^{b}_j.$$
The components of the inverse background metric denoted by $f^{ij}$ satisfy to the condition
$$f^{ik}f_{kj}=\delta^i_j.$$
Now, we can compare the background metric with the metric of gravitational field in every
point of the manifold $\Sigma_t$ in force of biectivity of mapping
$$\Sigma_t \longleftrightarrow {\cal T}(\Sigma_t)_{\bf x}.$$
The Levi--Civita connection $\bar\nabla_k$ is associated with the background metric $f_{ij}$:
$$\bar\nabla_k f_{ij}=0.$$

Let us define {\it scaled Dirac's conformal variables} $(\tilde\gamma_{ij}, \tilde\pi^{ij})$
by the following formulae:
\begin{equation}\label{generalized}
{{\tilde\gamma_{ij}:=\frac{\gamma_{ij}}{\sqrt[3]{\gamma /f}},\qquad
\tilde\pi^{ij}:=\sqrt[3]{\frac{\gamma}{f}}\left(\pi^{ij}-\frac{1}{3}\pi\gamma^{ij}\right)},}
\end{equation}
where additionally to the determinant $\gamma$ defined in (\ref{gammadet}),
the determinant of background metric $f$ is appeared:
$$f:=\det (f_{ij}).$$
The conformal metric $\tilde\gamma_{ij}$ (\ref{generalized}) is a tensor field, {\it id est}
it transforms according to the tensor representation of the group of diffeomorphisms.
The scaling variable $(\gamma/f)$ is a scalar field,
{\it id est} it is an invariant relative to diffeomorphisms.

We add to the conformal variables (\ref{generalized}) a canonical pair:
{\it a local intrinsic time} $D$ and a Hamiltonian density $\pi_D$
by the following way
\begin{equation}\label{DiracTpi}
{D:=-\frac{1}{3}\ln\left(\frac{\gamma}{f}\right),\qquad \pi_D:=\pi .}
\end{equation}
The formulae (\ref{generalized}), (\ref{DiracTpi}) define {\it the scaled Dirac's mapping} as a
mapping of the fiber bundles
\begin{equation}\label{generalizedD}
(\gamma_{ij}, \pi^{ij})\mapsto (D,\pi_D; \tilde\gamma_{ij}, \tilde\pi^{ij}).
\end{equation}
Riemannian superspaces of metrics $({}^3M)$ are defined on a compact Hausdorff manifolds $\Sigma_t$.
Denote the set, each point of which presents all various Riemannian metrics as ${\rm Riem}({}^3M)$.
Since the same Riemannian metric can be written in different coordinate systems,
we identify all the points associated with coordinate transformations of the diffeomorphism group
${\rm Diff}({}^3M)$.
All points receiving from some one by coordinate transformations of the group are called its orbit.
By this way the WDW superspace is defined as coset:
$$
{}^{(3)}\mathfrak{G}:={\rm Riem}({}^3M)/{\rm Diff} ({}^3M).
$$
As the saying goes \cite{Baierlein,Vision}, the superspace is the arena of Geometrodynamics.
Denote by ${}^{(3)}\mathfrak{G}{}^*$ the space of corresponding canonically conjugated densities of momenta.
According to the scaled Dirac's mapping (\ref{generalizedD}), we have the functional mapping of WDW phase
superspace of metrics $\gamma_{ij}$ and corresponding densities of their momenta $\pi^{ij}$
to WDW conformal superspace of metrics $\tilde\gamma_{ij}$, densities of momenta
$\tilde\pi^{ij}$; the local intrinsic time $D$, and the Hamiltonian density $\pi_D$:
$${{}^{(3)}\mathfrak{G}\times {}^{(3)}\mathfrak{G}{}^*
\to {}^{(3)}\mathfrak{\tilde{G}}\times {}^{(3)}\mathfrak{\tilde{G}}{}^*.}$$
We can conclude that the WDW phase superspace
$${}^{(3)}\mathfrak{\tilde{G}}\times {}^{(3)}\mathfrak{\tilde{G}}{}^*$$
is extended one, if we draw an analogy with relativistic mechanics \cite{Lanczos}.

York constructed the so-called extrinsic time \cite{York} as a trace of the extrinsic curvature tensor. Hence,
it is a scalar, and such a definition since that time is legalized in the theory of gravitation \cite{MTW}.
We built the intrinsic time with use of a ratio of the determinants of spatial metric tensors.
So to speak, the variables of the canonical pair: time - Hamiltonian density in the extended phase space,
in opposite to York's pair, are reversed.

Now the canonical variables are acquired a clear physical meaning.
The local time $D$ is constructed in accordance with the conditions imposed on the internal time.
The spatial metric $\gamma_{ij}$ carries information about the inner time.
According to the scaled Dirac's mapping (\ref{generalized}), (\ref{DiracTpi}),
the metric $\gamma_{ij}$ is factorized in the exponent function of the inner time $D$ and the conformal
metric $\tilde\gamma_{ij}$. So, the extracted intrinsic time has the spatial geometric origin.
Unlike to homogeneous examples, the intrinsic time $D$, in generic case, is local so-called many-fingered time.

The Poisson brackets (\ref{PB}) between new variables are the following
\begin{eqnarray}
\{D(t,{\bf x}), \pi_D(t,{\bf x}')\}&=&-\delta ({\bf x}-{\bf x}'),\nonumber\\
\{\tilde\gamma_{ij}(t,{\bf x}),\tilde\pi^{kl}(t,{\bf x}')\}&=&\tilde\delta_{ij}^{kl}\delta ({\bf x}-{\bf x}'),
\nonumber\\
\{\tilde\pi^{ij}(t,{\bf x}),\tilde\pi^{kl}(t,{\bf x}')\}&=&\frac{1}{3}(\tilde\gamma^{kl}\tilde\pi^{ij}-
\tilde\gamma^{ij}\tilde\pi^{kl}) \delta ({\bf x}-{\bf x}')\nonumber,
\end{eqnarray}
where
$$\tilde\delta_{ij}^{kl}:=\delta_i^k\delta_j^l+\delta_i^l\delta_j^k-
\frac{1}{3}\tilde\gamma^{kl}\tilde\gamma_{ij}$$
is the conformal Kronecker delta function  with properties:
$$\tilde\delta^{ij}_{ij}=5,~~~~~\quad \tilde\delta_{kl}^{ij}\tilde\delta_{mn}^{kl}=\tilde\delta_{mn}^{ij},\quad~~~~~
\tilde\delta_{ij}^{kl}\tilde\gamma_{kl}=\tilde\delta_{ij}^{kl}\tilde\gamma^{ij}=0,\quad~~~~~
\tilde\delta_{ij}^{kl}\tilde\pi^{ij}=\tilde\pi^{kl}.$$
The matrix $\tilde\gamma^{ij}$ is the inverse conformal metric, {\it id est}
$$\tilde\gamma^{ij}\tilde\gamma_{jk}=\delta^i_k,\qquad
\tilde\gamma^{ij}=\sqrt[3]{\frac{\gamma}{f}}\gamma^{ij}.$$
There are only five independent pairs $(\tilde\gamma_{ij}, \tilde\pi^{ij})$ per space point,
because of the properties
$$\tilde\gamma:=\det||\tilde\gamma_{ij}||=f,\qquad \tilde\pi:=\tilde\gamma_{ij}\tilde\pi^{ij}=0.$$
The generators $(D({\bf x},t)$, $\pi_D ({\bf x},t))$ form the subalgebra of the nonlinear Lie algebra.

Let us notice, that the problem of time and conserved dynamical quantities does not exist
in asymptotically flat worlds \cite{Regge}.
Time is measured by clocks of observers located at a sufficient far distance from the gravitational
objects. For this case the super-Hamiltonian (\ref{superHam}), constructed of the constraints,
is supplemented additionally by surface integrals at infinity.
Therefore, we focus our attention in the present paper on the cosmological problems only.

The Einstein's theory of gravitation is obtained from the shape dynamics by fixing the
St$\ddot{\rm u}$ckelberg's field (\ref{fixshape}):
\be
e^{-4\hat\phi}=\sqrt[3]{\frac{\gamma}{f}},
\ee
so that the factor is equal to
\be
\sqrt{\frac{f}{\gamma}}=\frac{e^{6\phi}}{<e^{6\phi}>_\gamma}.
\ee
The reparametrization constraints of shape dynamics are the following:
\be
{\cal H}=\int_\Sigma\,d^3x\sqrt\gamma\left(e^{6\hat\phi [\gamma, \pi, x)}-1\right),\quad
{\cal H}^i=-2\nabla_j\pi^{ij},\quad {\cal Q}=4(\pi-<\pi>\sqrt\gamma).\nonumber
\ee
Here $e^{6\hat\phi [\gamma, \pi, x)}$ is the solution of the Lichnerowicz -- York equation.

\section{Deparametrization}

For making deparametrization we utilize the substitution (\ref{generalized}) and obtain
\bea
\tilde\pi^{ij}\dot{\tilde\gamma}_{ij}&=&\left(\pi^{ij}-\frac{1}{3}\pi\gamma^{ij}\right)\dot\gamma_{ij}+
\sqrt[3]\gamma\left(\pi^{ij}-\frac{1}{3}\pi\gamma^{ij}\right)\gamma_{ij}\left(\gamma^{-1/3}\right)^{.}=\nonumber\\
&=&\pi^{ij}\dot\gamma_{ij}-\frac{1}{3}\pi\left(\ln\gamma\right)^{.}.\nonumber
\eea
Then, after using the variables (\ref{DiracTpi}), we get
$$
\pi^{ij}\frac{d}{dt}\gamma_{ij}=\tilde\pi^{ij}\frac{d}{dt}\tilde\gamma_{ij}-\pi_D\frac{d}{dt}D.
$$
Then ADM functional of action (\ref{ADMvar}) takes the form
\be
W=\int\limits_{t_I}^{t_0} dt\int\limits_{\Sigma_t} d^3x\left[
\left(\tilde\pi^{ij}_L+\tilde\pi^{ij}_{TT}\right)\frac{d}{dt}\tilde\gamma_{ij}-
\pi_D\frac{d}{dt}D-
N{\cal H}_\bot-N^i{\cal H}_i\right],\label{actionHilbert}
\ee
where the conformal momentum densities are decomposed on longitudinal and traceless - transverse parts,
the trace of momentum is expressed through the scalar of extrinsic curvature:
$$
\tilde\pi^{ij}:=\tilde\pi^{ij}_L+\tilde\pi^{ij}_{TT},\qquad K=\frac{\pi}{2\sqrt\gamma}.
$$
The momentum $\pi_D$ is expressed out of the Hamiltonian constraint (\ref{H})
\be\label{pfound}
\pi_D [\tilde\pi^{ij}_L, \tilde\pi^{ij}_{TT},\tilde\gamma_{ij}, D]=
\sqrt{6\gamma}
\left[8\Psi^5\tilde\Delta\Psi
-\tilde{R}\Psi^4+\tilde\pi_{ij}\tilde\pi^{ij}+
\tilde{T}_{\bot\bot}\right]^{1/2},
\ee
where
$$\Psi=\left(\frac{\gamma}{f}\right)^{1/12}.$$
The property
$$
\tilde{A}_{ij}\tilde{A}^{ij}=\tilde\pi_{ij}\tilde\pi^{ij}
$$
was utilized here.

Picking $D$ to be the time and setting an equality of the time intervals
$$dD=dt,$$
we can partially reduce the action (\ref{actionHilbert}).

Substituting the expressed $\pi_D$ (\ref{pfound})
into the ADM functional of action (\ref{actionHilbert}),
we get the functional presimplectic 1-form \cite{Olver,PTwo}  with local time $D$
\be
\omega^1=\int\limits_{\Sigma_D}
d^3x\left[\left(\tilde\pi^{ij}_L+\tilde\pi^{ij}_{TT}\right)
d\tilde\gamma_{ij}-
\pi_D
\left(\tilde\pi^{ij}_L,\tilde\pi^{ij}_{TT},\tilde\gamma_{ij}, D\right)d D
\right].\label{omega1}
\ee

\section{Global time}

A global time exists in homogeneous cosmological models (see, for example, papers
\cite{Kasner,Misner,Kuchar,Isham,Pavlov,PavlovFlat,Pavlov1}).
For getting a global time in general  case, let us set the CMC gauge \cite{York}
(consideration especially for closed manifolds see \cite{JamesI}) on every slice, which labeled by the
coordinate time $t$,
\be\label{restrictions}
K\equiv -3\kappa=K(t),
\ee
where
$$\kappa:=\frac{1}{3}(\kappa_1+\kappa_2+\kappa_3)$$
is a mean curvature of the hypersurface $\Sigma_t$ -- arithmetic mean of the principal curvatures
$\kappa_1, \kappa_2, \kappa_3$.

Instead of Dirac's variables $(D, \pi_D)$ (\ref{hatTpT}) for obtaining global canonical conjugated variables
$$
\int_\Sigma\,d^3y D(y),\qquad
\int_\Sigma\,d^3x \pi_D (x)
$$
it is preferable to take York's transparent ones
\be\label{York}
T=-\sqrt\gamma,\qquad \pi_T=\frac{2\pi}{3\sqrt\gamma}.
\ee
Although, $T$ is a scalar density, we need not construct a scalar to get a global time.
Notice, that in generic case, a tensor density of weight $n$ is such a quantity $\tau$, that
$$\tau=\gamma^{n/2}T,$$
where $T$ is a tensor field.

A volume of a hypersurface
$$V:=\int_\Sigma d^3x\sqrt\gamma $$
plays a role of global time and a mean value of density of momentum
\be\label{reducedHam}
H :=\frac{2}{3}<\pi>=\frac{2}{3}\frac{\int_\Sigma d^3x \pi (x)}{\int_\Sigma d^3x\sqrt{\gamma (x)}}.
\ee
is a Hamiltonian.
Let us find the Poisson bracket between these nonlocal characteristics
$$
\left\{
\int_\Sigma\,d^3y\sqrt\gamma (y),\frac{2}{3}\frac{\int_\Sigma\,d^3y\pi (y)}{\int_\Sigma\,d^3y\sqrt\gamma (y)}
\right\}.
$$
We calculate functional derivatives of the functionals defined above
\bea
\frac{\delta}{\delta\gamma_{ij}(x)}{\int_\Sigma\,d^3y\sqrt\gamma (y)}&=&
\frac{1}{2}\sqrt\gamma (x)\gamma^{ij}(x);\nonumber\\
\frac{\delta}{\delta\pi^{ij}(x)}<\pi>&=&
\frac{1}{V}\frac{\delta}{\delta\pi^{ij}(x)}\int_\Sigma\,d^3y\pi (y)=\nonumber\\
&=&\frac{1}{V}\frac{\delta}{\delta\pi^{ij}(x)}\int_\Sigma\,d^3y\pi^{ij}(y)\gamma_{ij}(y)=
\frac{1}{V}\gamma_{ij}(x).\nonumber
\eea
Hence, a canonical pair is emerged:
\be
\{V, \frac{2}{3}<\pi>\}=\frac{1}{2V}\int_\Sigma\,d^3x\sqrt\gamma(x)\gamma^{ij}(x)\gamma_{ij}(x)=1.\nonumber
\ee
The corresponding Poisson bracket is canonical as in Ashtekar's approach \cite{Gorobey}.
Then the second term in (\ref{actionHilbert}) can be processed because of
$$
\frac{dD}{dt}=-\frac{1}{3\gamma}\frac{d\gamma}{dt}=-\frac{2}{3\sqrt\gamma}\frac{d\sqrt\gamma}{dt}.
$$
The York's condition (\ref{restrictions}) sets a slicing,
allowing to obtain a global time.
After Hamiltonian reduction and deparametrization procedures were executed, we yield the action
\be\label{actionreduction}
W=\int\limits_{V_I}^{V_0} dV\int\limits_{\Sigma_t} d^3x\left[
\tilde\pi^{ij}\frac{d\tilde\gamma_{ij}}{d V}\right]-
\int\limits_{V_I}^{V_0} H\, d V
-\int\limits_{V_I}^{V_0}\, dV\int\limits_{\Sigma_t} d^3x N^i {\cal H}_i
\ee
with the Hamiltonian (\ref{reducedHam}).

The Hamiltonian constraint is algebraic of the second order relative to $K$ that is
characteristic for relativistic theories. The coordinate time $t$ parameterizes the constrained theory.
The Hamiltonian constraint is a result of a gauge arbitrariness of the spacetime slicing
into space and time.

Let us notice, if the York's time was chosen \cite{KKuchar, Isenberg},
one should have to resolve the Hamiltonian constraint with respect to the variable $D$,
that looks unnatural difficult.
As was noted in \cite{Wald}, if the general relativity could be deparametrized,
a notion of total energy in a closed universe could well emerge. As follows from (\ref{pfound}),
in a generic case, the energy of the Universe is not conserved.

In CMC gauge we one yields
\be
<\pi>=\frac{2}{V}\int_\Sigma\, d^3x\sqrt\gamma K=2K.
\ee
The CMC gauge make condition to the lapse function $N$, if suppose the shift vector is zero $N_i=0$:
$$
K_{ij}=-\frac{1}{2N}\dot\gamma_{ij}.
$$
Then the trace is
$$K=\gamma^{ij}K_{ij}=-\frac{1}{2N}\gamma^{ij}\dot\gamma_{ij}=-\frac{1}{2N}\left(\ln\gamma\right)^{.}.$$

So, we find that the lapse function is not arbitrary, but defined according to the following restriction
\be
N=-\frac{1}{2K}\frac{d}{dt}\ln\gamma .
\ee

In a particular case, when $K={\rm const}$  for every hypersurface,
the Hamiltonian does not depend on time, so the energy of the system is conserved.
One maximal slice $(K=0)$ exists at the moment of time symmetry.
If additionally we were restricted our consideration conformal flat spaces, we should got set of waveless
solutions \cite{Isenberg}. Black holes, worm holes belong to this class of solution.

\section{Global intrinsic time in FRW universe}

The observational Universe with high precision is homogeneous and isotropic \cite{Planck}.
So, it is natural to choose a compact conformally flat space of modern scale $a_0$
as a background space. It is necessary to study an applicability of the intrinsic global time chosen to nearest non-symmetric cases by
taking into account linear metric perturbations.
The first quadratic form
$${\bf f}=f_{ij}({\bf x})dx^i\otimes dx^j$$
in spherical coordinates  $\chi, \theta, \varphi$
\begin{equation}\label{back}
a_0^2\left[d\chi^2+\sin^2\chi (d\theta^2+\sin^2\theta d\varphi^2)\right]\equiv
a_0^2({}^1f_{ij})dx^i dx^j
\end{equation}
should be chosen for a background space of positive curvature.
We introduced the metric of invariant space $({}^1f_{ij})$ with a curvature ${\cal K}=\pm 1, 0$.
For a pseudosphere in (\ref{back}) instead of $\sin\chi$ one should take $\sinh\chi$, and for a flat space
one should take $\chi.$ The spatial metric is presented as
\begin{equation}\label{Fmetric}
(\gamma_{ij})=a^2(t)({}^1f_{ij})=e^{-2D}(f_{ij})=\left(\frac{a}{a_0}\right)^2\tilde\gamma_{ij}.
\end{equation}
Hence, one yields an equality:
$\tilde\gamma_{ij}=f_{ij},$
so, in our high symmetric case, the conformal metric is equal to the background one.

The energy-momentum tensor is of the form
$$
(T_{\mu\nu})=\left(
\begin{array}{cc}
N^2 \rho&0\\
0&p\gamma_{ij}
\end{array}
\right),
$$
where $\rho$ is a background energy density of matter, and $p$ is its pressure.
We have for the matter contribution
\begin{equation}\label{matterT}
\sqrt\gamma T_{\bot\bot}=\sqrt{({}^1 f)}a^3\rho ,
\end{equation}
because of
$$T_{\bot\bot}=n^\mu n^\nu T_{\mu\nu}=\frac{1}{N^2}T_{00}=
\rho,\qquad n^\mu=\left(\frac{1}{N},-\frac{N^i}{N}\right).$$

The functional of action in the symmetric case considered here takes the form
\begin{equation}\label{Wh}
W^{(0)}=-\int\limits_{t_I}^{t_0}\, dt\int\limits_{\Sigma_t}\, d^3x
\pi_D\frac{d D}{dt}-\int\limits_{t_I}^{t_0}\, dt\int\limits_{\Sigma_t}\, d^3x
N{\cal H}_\bot .
\end{equation}
Let us present results of necessary calculations been executed
\begin{eqnarray}
K_{ij}&=&-\frac{1}{2N}\frac{d\gamma_{ij}}{dt}=\frac{1}{N}\frac{d D}{dt}\gamma_{ij},\nonumber\\
K&=&\gamma^{ij}K_{ij}=\frac{3}{N}\frac{d D}{dt},\nonumber\\
\left(K_{ij}K^{ij}-K^2\right)&=&-\frac{6}{N^2}\left(\frac{d D}{dt}\right)^2,\nonumber\\
\pi_D\frac{d D}{dt}&=&
4\sqrt\gamma K\frac{d D}{dt}=\frac{12}{N}\sqrt{f}e^{-3D}\left(\frac{d D}{dt}\right)^2,\nonumber\\
R&=&\frac{6 {\cal K}}{a_0^2}e^{2D}=\frac{6{\cal K}}{a^2}.\nonumber
\end{eqnarray}
The Hamiltonian constraint
\begin{equation}\label{Ham0Class}
{\cal H}_\bot=\sqrt{\gamma}\left(K_{ij}K^{ij}-K^2\right)-\sqrt{\gamma}R+\sqrt{\gamma} T_{\bot\bot},
\end{equation}
takes the following form
$$
{\cal H}_\bot=-6\sqrt{f}e^{-3D}\left[\frac{1}{N^2}
\left(\frac{dD}{dt}\right)^2+\frac{\cal K}{a_0^2}e^{2D}-\frac{1}{6}\rho \right].
$$

The action (\ref{Wh}) presents a model of the classical mechanics \cite{PP}
\begin{equation}\label{Whint}
W^{(0)}=V_0\int\limits_{t_I}^{t_0}\, dt
\left[-\frac{6}{N}e^{-3D}\left(\frac{d D}{dt}\right)^2+\frac{6{\cal K}}{a_0^2}N e^{-D}-
N e^{-3D}\rho \right]
\end{equation}
after integration over the slice $\Sigma_0$, where
\begin{equation}\label{Sigma0}
V_0:=\int\limits_{\Sigma_0}\sqrt{f}\, d^3x
\end{equation}
is a volume of the space with a scale $a_0$. We rewrite it in Hamiltonian form
\begin{equation}\label{LG}
W^{(0)}=\int\limits_{t_I}^{t_0}\, dt\left[p_D\frac{d D}{dt}-N H_{\bot}\right],
\end{equation}
where
\begin{equation}\label{clHam}
H_{\bot}=-\frac{1}{24V_0}e^{3D}p_D^2-\frac{6{\cal K}V_0}{a_0^2}e^{-D}+V_0e^{-3D}\rho
\end{equation}
is the Hamiltonian constraint as in classical mechanics.
Here $D(t)$ is a generalized coordinate, and $p_D$ is its canonically conjugated momentum
\begin{equation}\label{pa}
p_D=\frac{\delta W^{(0)}}{\delta\dot D} =-\frac{12 V_0}{N}e^{-3D}\left(\frac{d D}{dt}\right).
\end{equation}
Resolving the constraint (\ref{clHam}), we get the energy
\begin{equation}\label{paE}
p_D^2=E^2 (D),\qquad E(D):= 2\sqrt{6}V_0 e^{-2D}\sqrt{e^{-2D}\rho -\frac{6{\cal K}}{a_0^2}},
\end{equation}
that was lost in Standard cosmology \cite{MY}.

The equation of motion is obtained from the action (\ref{LG}):
\begin{equation}\label{eqnmotion}
\dot{p}_D=N\left[
\frac{e^{3D}}{8V_0}p_D^2-\frac{6{\cal K}V_0}{a_0^2}e^{-D}+3V_0e^{-3D}\rho-V_0e^{-3D}
\frac{\partial{\rho}}{\partial D}
\right] .
\end{equation}

From the Hamiltonian constraint (\ref{clHam}), the equation of motion (\ref{eqnmotion}), and the energy
continuity equation
\begin{equation}\label{eceqn}
\dot\rho=-3(\rho+p)\frac{\dot{a}}{a},
\end{equation}
we get Einstein's equations in standard form:
\begin{eqnarray}
\left(\frac{\dot{a}}{a}\right)^2+\frac{\cal{K}}{a^2}&=&\frac{1}{6}\rho,\nonumber\\
2\frac{\ddot{a}}{a}+\left(\frac{\dot{a}}{a}\right)^2+\frac{\cal{K}}{a^2}&=&-\frac{1}{2}p.\nonumber
\end{eqnarray}
Here we have used the normal slicing condition $N=1$.

Substituting $p_D$ from (\ref{pa}) into (\ref{paE}), we obtain the famous {\it Friedmann equation}
\begin{equation}\label{Friedmann}
\frac{1}{N^2 e^{2D}}{\left(\frac{d D}{dt}\right)}^2+\frac{\cal K}{a_0^2}=\frac{1}{6}e^{-2D}\rho .
\end{equation}
From the Standard cosmology point of view, it connects the expansion rate of the Universe (Hubble parameter)
\begin{equation}\label{Hubblep}
H:=\left(\frac{\dot{a}}{a}\right)=-\frac{d D}{dt}
\end{equation}
with an energy density of matter $\rho$ and a spatial curvature
\begin{equation}\label{Hubble0}
{H^2\equiv \left(\frac{d D}{dt}\right)^2= \frac{1}{6}
\left(\rho_M+\rho_{rigid}+\rho_{rad}+\rho_{curv}\right).}
\end{equation}
In the right side of equation (\ref{Hubble0}),
$\rho_M$ is an energy density of nonrelativistic matter
$$\rho_M=\rho_{M,0}\left(\frac{a_0}{a}\right)^3,$$
$\rho_{rigid}$ is an energy density of matter
$$\rho_{rigid}=\rho_{rigid,0}\left(\frac{a_0}{a}\right)^6,$$
with a rigid state equation \cite{Zeld}
$${p=\rho,}$$
$\rho_{rad}$ is an energy density of radiation
$$\rho_{rad}=\rho_{rad,0}\left(\frac{a_0}{a}\right)^4,$$
$\rho_{curv}$, which is defined as
$$\frac{1}{6}\rho_{curv}:= -\frac{\cal K}{a_0^2}e^{2D}=-\frac{\cal K}{a^2},$$
is a contribution from the spatial curvature. In the above formulae
$\rho_{M,0};$ $\rho_{rigid,0};$ $\rho_{rad,0};$ $\rho_{curv,0}$ are modern values of the densities.
Subsequently, the equation for the energy (\ref{paE}) can be
rewritten as
$$E(D)=2\sqrt{6}V_0e^{-3D}\sqrt{\rho+\rho_{curv}}.$$
The $CDM$ model considered has not dynamical degrees of freedom.
{\it According to the Conformal cosmology interpretation,
the Friedmann equation (\ref{Hubble0}) has a following sense:
it ties the intrinsic time interval $dD$ with the coordinate time
interval $dt$, or the conformal interval $d\eta=dt/a$.}
Let us note, that in the left side of the Friedmann equation (\ref{Hubble0}) we see the square of the
extrinsic York's time which is proportional to the square of the Hubble parameter. If we choose the
extrinsic time, the equation (\ref{Hubble0}) becomes algebraic of the second order,
and the connection between temporal intervals should be
lost\footnote{``The time is out of joint''. William Shakespeare: {\it Hamlet}. Act 1. Scene V. Longman,
London (1970).}.

In an observational cosmology, the density can be expressed in terms of the present-day
critical density $\rho_{c}$:
$$
\rho_c (a)\equiv 6H_0^2,
$$
where
$H_0$ is a modern value of the Hubble parameter. Further, it is convenient to use density parameters as ratio
of present-day densities
$$
\Omega_{M}\equiv\frac{\rho_{M,0}}{\rho_c},\quad \Omega_{rigid}\equiv\frac{\rho_{rigid,0}}{\rho_c},\quad
\Omega_{rad}\equiv\frac{\rho_{rad,0}}{\rho_c},\quad
\Omega_{curv}\equiv\frac{\rho_{curv,0}}{\rho_c},
$$
satisfying the condition
$$
\Omega_{M}+\Omega_{rigid}+\Omega_{rad}+\Omega_{curv}=1.
$$
Then, one yields
\begin{eqnarray}\nonumber
H^2&=&\frac{1}{6}\rho_c\left[\Omega_{M}\left(\frac{a_0}{a}\right)^3+\Omega_{rigid}
\left(\frac{a_0}{a}\right)^6+
\Omega_{rad}\left(\frac{a_0}{a}\right)^4+
\Omega_{curv}\left(\frac{a_0}{a}\right)^2\right].\nonumber
\end{eqnarray}
According to NASA diagram, 25\% of the Universe is dark matter,
70\% of the Universe is dark energy about which practically nothing is known.

After transition to conformal variables
\begin{equation}\label{confvar}
Ndt=a_0e^{-D}d\eta,\qquad \tilde\rho=e^{-4D}\rho ,
\end{equation}
the Friedmann equation takes the form
\begin{equation}
\left(\frac{d D}{d\eta}\right)^2+{\cal K}=\frac{1}{6}a_0 N^2 e^{2D}\tilde\rho.
\end{equation}

\section{Global time in perturbed FRW universe}

Let us consider additional corrections to the Friedmann equation from non ideal FRW model, taking into account
metric perturbations.
The metric of perturbed FRW universe can be  presented as
\begin{equation}\label{gmnperturb}
g_{\mu\nu}=a^2(t)\left({}^1 f_{\mu\nu}+h_{\mu\nu}\right),
\end{equation}
where $({}^1 f_{\mu\nu})$ is the metric of spacetime with the spatial metric $({}^1 f_{ij})$ considered above,
deviations $h_{\mu\nu}$ are assumed small.
The perturbation $h_{\mu\nu}$ is not a tensor in the perturbed universe, nonetheless we define
$$
h_\mu^\nu\equiv ({}^1 f^{\nu\rho})h_{\rho\mu},\qquad
h^{\mu\nu}\equiv({}^1 f^{\mu\rho})({}^1 f^{\nu\sigma})h_{\rho\sigma}.
$$

For a coordinate system chosen in the background space, there are various possible coordinate systems in the
perturbed spacetime. In GR perturbation theory, {\it a gauge transformation} means a coordinate transformation
between coordinate systems in the perturbed spacetime.  The coordinates of the background spacetime are kept
fixed, the correspondence between the points in the background and perturbed spacetime is changing.
A manifestly gauge invariant cosmological perturbation theory was built by
James Bardeen \cite{Bardeen}, and analyzed in details by Hideo Kodama and Misao Sasaki \cite{Kodama}.
Now, keeping the gauge chosen, {\it id est} the correspondence between the background and perturbed spacetime
points, we implement coordinate transformation in the background spacetime. Because of our background coordinate
system was chosen to respect the symmetries of the background, we do not want to change our slicing.
Eugene Lifshitz made decomposition of perturbations of metric and energy -- momentum tensor into scalar, vortex,
and tensor contributions refer to their transformation properties under rotations in the background space
in his pioneer paper \cite{Lifshitz}.
The scalar perturbations couple to density and pressure perturbations. Vector perturbations couple to rotational
velocity perturbations. Tensor perturbations are nothing but gravitational waves.

$\bullet$
In a flat case (${\cal K}=0$) the eigenfunctions of the Laplace -- Beltrami operator are plane waves.
For arbitrary perturbation $f(t, {\bf x})$ we can make an expansion
$$f(t, {\bf x})=\sum\limits_{{\bf k}=0}^\infty f_{\bf k}(t)e^{\imath {\bf k}\cdot{\bf x}}$$
over Fourier modes. We can consider them in future as a particular case of models with constant curvature.
Let us proceed the harmonic analysis of linear geometric perturbations using irreducible representations of
isometry group of the corresponding constant curvature space \cite{Durrer,Ruth}.

$\bullet$ Scalar harmonic functions.

Let us define for a space of positive curvature ${\cal K}=1$ an invariant space with the first quadratic form
\begin{equation}\label{inv}
({}^1 f_{ij})dx^i dx^j=d\chi^2+\sin^2\chi (d\theta^2+\sin^2\theta d\varphi^2).
\end{equation}
The eigenfunctions of the Laplace -- Beltrami operator form a basis of unitary representations of the group of
isometries of the three-dimensional space $\Sigma$ (\ref{inv}) of constant unit curvature.
In particular, the eigenfunctions on a sphere $S^3$ are the following \cite{GMM}:
\begin{equation}\label{eigenfunctionssphere}
Y_{\lambda l m} (\chi, \theta, \varphi)=\frac{1}{\sqrt{\sin \chi}}
\sqrt{\frac{\lambda (\lambda+l)!}{(\lambda-l+1)!}}
P_{\lambda-1/2}^{-l-1/2}(\cos \chi)Y_{lm}(\theta, \varphi).
\end{equation}
Here, $P_\mu^\nu (z)$ are attached Legendre functions, $Y_{lm}(\theta, \varphi)$ are spherical functions,
indices run the following values
$$\lambda=1,2,\ldots;\qquad l=0,1,\ldots,\lambda-1; \qquad m=-l,-l+1,\ldots, l.$$
There is a condition of orthogonality and normalization for functions  (\ref{eigenfunctionssphere})
\begin{equation}\label{ort}
\int\limits_0^\pi d\chi\sin^2\chi\int\limits_0^\pi
d\theta\sin\theta\int\limits_0^{2\pi} d\varphi\,Y_{\lambda l m}^{*}(\chi, \theta, \varphi)
Y_{\lambda' l' m'}(\chi, \theta, \varphi)=\delta_{\lambda\lambda'}\delta_{ll'}\delta_{mm'}.
\end{equation}
In a hyperbolic case with negative curvature ${\cal K}=-1$ the eigenfunctions are following \cite{GMM}:
\begin{equation}\label{eigenfunctionspseudosphere}
Y_{\lambda l m} (\chi, \theta, \varphi)=\frac{1}{\sqrt{\sinh \chi}}
\frac{|\Gamma(\imath\lambda+l+1)|}{|\Gamma (\imath\lambda)|}
P_{\imath\lambda-1/2}^{-l-1/2}(\cosh \chi)Y_{lm}(\theta, \varphi),
\end{equation}
where $0\le\lambda <\infty; l=0,1,2,\ldots; m=-l,-l+1,\ldots, l$.
There is a condition of orthogonality and normalization for functions  (\ref{eigenfunctionspseudosphere})
$$\int\limits_0^\Lambda d\chi\sinh^2\chi\int\limits_0^\pi d\theta\sin\theta\int\limits_0^{2\pi}
d\varphi\,Y_{\lambda l m}^{*}(\chi, \theta, \varphi)Y_{\lambda' l' m'}(\chi, \theta, \varphi)=
\delta (\lambda-\lambda')\delta_{ll'}\delta_{mm'},$$
where $\Lambda$ is some cut-off limit.

The equation on eigenvalues can be presented in the following symbolic form
\begin{equation}\label{LaplaceB}
(({}^1\bar{\rm\Delta})+k^2)Y_{\bf k}^{(s)}=0,
\end{equation}
where $-k^2$ is an eigenvalue of the Laplace -- Beltrami operator $({}^1\bar{\rm\Delta})$ on $\Sigma$.
The connection $({}^1\bar\nabla)$ is associated with the metric of invariant space (\ref{inv}).
For a flat space $({\cal K}=0)$ the eigenvectors of the equation (\ref{LaplaceB})
are flat waves as the unitary irreducible representations of the Euclidean translation group.
For a positive curvature space $({\cal K}>0)$  we have $k^2=l(l+2)$, and for a space of negative curvature
$({\cal K}<0)$ we have $k^2>1$.
The scalar contributions of the vector and the symmetric tensor fields can be expanded in terms of
\begin{eqnarray}
Y_{{\bf k},i}^{(s)}:&=&-\frac{1}{k}({}^1\bar\nabla_i) Y_{\bf k}^{(s)},\label{pr1}\\
Y_{{\bf k},{ij}}^{(s)}:&=&\frac{1}{k^2}({}^1\bar\nabla_i) ({}^1\bar\nabla_j) Y_{\bf k}^{(s)}+
\frac{1}{3}({}^1 f_{ij})Y_{\bf k}^{(s)}\label{pr2} .
\end{eqnarray}
Different modes do not couple in the linearized approximation.
So we are able to consider a contribution of a generic mode.

Let us start with considering scalar modes because of their main contribution to galaxy formation. Linear
perturbations of the four-metric in terms of the Bardeen gauge invariant potentials $\rm\bar\Psi (t)$ and
$\rm\Phi (t)$ in conformal--Newtonian gauge are of the
\begin{equation}\label{gmn}
ds^2=-N_D^2\left(1+\rm\bar\Psi Y_{\bf k}^{(s)}\right)^2dt^2+
a^2\left(1-\rm\Phi Y_{\bf k}^{(s)}\right)^2 ({}^1 f_{ij}) dx^i dx^j,
\end{equation}
where a summation symbol over harmonics is omit, $N_D$ is the Dirac's lapse function defined as
\begin{equation}\label{NDirac}
N\equiv N_D \left(1+\rm\bar\Psi Y_{\bf k}^{(s)}\right).
\end{equation}
The spatial metric is presented as a sum of the Friedmann metric (\ref{Fmetric})
considered in the previous section and a perturbed part
\begin{equation}\label{Fmetricp}
(\gamma_{ij})=a^2(t)({}^1 f_{ij})-2a^2(t){\rm\Phi} Y_{\bf k}^{(s)}({}^1 f_{ij}).
\end{equation}
The determinant of the spatial metric (\ref{Fmetricp}) in the first order of accuracy is
\begin{equation}\label{gfij}
{\det} (\gamma_{ij})=a^6(t)\left(1-6\rm\Phi Y_{\bf k}^{(s)}\right){\det} ({}^1 f_{ij}).
\end{equation}
For the high symmetric Friedmann case, the background metric (\ref{back}) coincides with the conformal one:
$f_{ij}=\tilde\gamma_{ij}$
because of
\begin{equation}\label{gammafij}
\left(\frac{\gamma}{f}\right)^{1/3}=\left(\frac{a(t)}{a_0}\right)^2
\left(1-2\rm\Phi Y_{\bf k}^{(s)}\right)=e^{-2D(t)}\left(1-2\rm\Phi Y_{\bf k}^{(s)}\right).
\end{equation}
One obtains from relations (\ref{Fmetricp}), (\ref{gammafij}) the connection between components of perturbed
metric tensor and the background one
\begin{equation}\label{gammaijfij}
(\gamma_{ij})=e^{-2D(t)}\left(1-2\rm\Phi Y_{\bf k}^{(s)}\right)(f_{ij}),\qquad
(\gamma^{ij})=e^{2D(t)}\left(1+2\rm\Phi Y_{\bf k}^{(s)}\right)(f^{ij}).
\end{equation}

Now, we calculate components of the tensor of extrinsic curvature
\begin{equation}\label{Kzeroth}
K_{ij}=\frac{1}{N_D}\left[\frac{d D}{dt}-\left(({\rm\bar\Psi}+2{\rm\Phi})\frac{d D}{dt}-
\frac{d\rm\Phi}{dt}\right)Y_{\bf k}^{(s)}\right]\gamma_{ij},
\end{equation}
and its trace
\begin{equation}
K=\gamma^{ij}K_{ij}=\frac{3}{N_D}\left[\frac{d D}{dt}-\left(({\rm\bar\Psi}+2{\rm\Phi})
\frac{d D}{dt}-\frac{d\rm\Phi}{dt}\right)
Y_{\bf k}^{(s)}\right].
\end{equation}
Then we obtain
\begin{equation}
\nonumber
(K_{ij}K^{ij}-K^2)=-\frac{6}{N_D^2}\left[
\left(\frac{d D}{dt}\right)^2-2\left(\frac{d D}{dt}\right)
\left(({\rm\bar\Psi}+2{\rm\Phi})\frac{d D}{dt}-\frac{d\rm\Phi}{dt}\right)Y_{\bf k}^{(s)}
\right].
\end{equation}
From the relation between metric tensors (\ref{gammaijfij}) one obtains the relation between the
corresponding Jacobians
$$\sqrt\gamma=e^{-3D(t)}\left(1-3\rm\Phi Y_{\bf k}^{(s)}\right)\sqrt{f}.$$
The calculation gives
\bea
&&\pi_D\frac{d D}{dt}=4\sqrt{\gamma}K\frac{d D}{dt}=\nonumber\\
&=&\frac{12}{N_D}e^{-3D}\sqrt{f}
\left(\frac{d D}{dt}\right)^2-
\frac{12}{N_D}e^{-3D}\sqrt{f}\frac{d D}{dt}
\left(({\rm\bar\Psi}+5{\rm\Phi})\frac{d D}{dt}-\frac{d\rm\Phi}{dt}\right)
Y_{\bf k}^{(s)}.\nonumber
\eea
We need to the perturbation of the Ricci scalar additionally.
According to the Palatini identity \cite{Lifshitz} from the differential geometry, a variation of the Ricci
tensor is set by the formula
\begin{equation}\label{Palatini}
\delta R_{ij}=\frac{1}{2}\left(\nabla_n\nabla_j\delta\gamma_i^n+
\nabla_n\nabla_i\delta \gamma_j^n-\nabla_j\nabla_i\delta \gamma_n^n-
{\rm\Delta}\delta\gamma_{ij}\right).
\end{equation}
Here the Levi--Civita connection $\nabla_i$ is associated with the Friedmann metric $\gamma_{ij}$
(\ref{Fmetric}).
The metric variations are expressed through harmonics (\ref{Fmetricp})
\begin{equation}\label{deltagammaij}
\delta\gamma_{ij} = -2a^2(t){\rm\Phi} Y_{\bf k}^{(s)}({}^1 f_{ij}),
\end{equation}
the indices of the metric variations are moved up with respect to the background metric
\begin{equation}
\delta\gamma_i^n=\gamma^{nj}\delta\gamma_{ji}=-2{\rm\Phi} Y_{\bf k}^{(s)}\delta_i^n,\qquad
\delta\gamma_n^n=-6{\rm\Phi} Y_{\bf k}^{(s)}.
\end{equation}
Substituting them into the (\ref{Palatini}), one obtains
$$
\delta R_{ij}={\rm\Phi}\left[(\nabla_j\nabla_i-\nabla_i\nabla_j)+\nabla_j\nabla_i+
\gamma_{ij}{\rm\Delta}\right]Y_{\bf k}^{(s)}.
$$
Using the Laplace -- Beltrami equation (\ref{LaplaceB}), commutativity of the covariant differentiation
operators, and properties of harmonics (\ref{pr1}), (\ref{pr2}), one gets
\begin{equation}
\delta R_{ij}={\rm\Phi} k^2\left[Y_{{\bf k},{i j}}^{(s)}-\frac{4}{3} \left({}^1 f_{ij}\right)
Y_{\bf k}^{(s)}\right].
\end{equation}
Remark the connection between the operators, used above
$\left(\rm\Delta\right)=\left({}^1\bar{\rm\Delta}\right)/{a^2}.$
For getting the variation of the Ricci scalar we make summation
$$\delta R=\gamma^{ij}\delta R_{ij}=-\frac{4}{a^2}{\rm\Phi} k^2 Y_{\bf k}^{(s)},$$
where we used the property of traceless of the tensor
$Y_{{\bf k},i}^{i(s)}=0.$
Finally, we yield
\begin{equation}\label{Ricciscalar}
\sqrt\gamma R=\frac{6{\cal K}}{a_0^2}\sqrt{f}e^{-D(t)}-
\frac{4}{a_0^2}\sqrt{f}e^{-D(t)}{\rm\Phi} k^2 Y_{\bf k}^{(s)}.
\end{equation}

Let us consider the first order corrections to terms of the action (\ref{Wh}),
taking into account the connection between lapse functions (\ref{NDirac}),
\begin{eqnarray}
&&N\sqrt\gamma\left(K_{ij}K^{ij}-K^2\right)=-\frac{6}{N_D}e^{-3D}\sqrt{f}\left(\frac{d D}{dt}\right)^2-
\label{piDfirst}\\
&-&\frac{6}{N_D}e^{-3D}\sqrt{f}\left[-({\rm\bar\Psi} + 7{\rm\Phi})\left(\frac{d D}{dt}\right)^2+
2\left(\frac{d D}{dt}\right)
\left(\frac{d{\rm\Phi}}{dt}\right)\right]Y_{\bf k}^{(s)}.\nonumber
\end{eqnarray}
The curvature term after the correction is the following
\begin{equation}
N\sqrt\gamma R=\frac{2}{a_0^2}N_D\sqrt{f}e^{-D}\left(3{\cal K}+
(3{\cal K}{\rm\bar\Psi}-2k^2{\rm\Phi})Y_{\bf k}^{(s)}\right).
\end{equation}
The matter term with the correction has a form
\begin{equation}
N\sqrt\gamma T_{\bot\bot}=N_De^{-3D}\sqrt{f}\rho\left(1+(\rm\bar\Psi-3\rm\Phi)Y_{\bf k}^{(s)}\right).
\end{equation}
The Hamiltonian constraint, multiplied to the lapse function, is the following
\begin{eqnarray}
N{\cal H}_\bot&=&-\frac{6}{N_D}e^{-3D}\sqrt{f}\left(\frac{d D}{dt}\right)^2-
\frac{6{\cal K}}{a_0^2}N_De^{-D}\sqrt{f}+N_D e^{-3D}\sqrt{f}\rho+\nonumber\\
&+&\frac{6}{N_D}e^{-3D}\sqrt{f}\left[({\rm\bar\Psi}+7{\rm\Phi})\left(\frac{d D}{dt}\right)^2-
2\left(\frac{d D}{dt}\right)
\left(\frac{d{\rm\Phi}}{dt}\right)\right]Y_{\bf k}^{(s)}-\nonumber\\
&-&\frac{2}{a_0^2}N_De^{-D}\sqrt{f}(3{\cal K}{\rm\bar\Psi}-2k^2{\rm\Phi})Y_{\bf k}^{(s)}+
N_De^{-3D}\sqrt{f}\rho ({\rm\bar\Psi}-3{\rm\Phi}))Y_{\bf k}^{(s)}.\nonumber
\end{eqnarray}
For the first order correction to the Lagrangian we get the following expression
\begin{eqnarray}
&&\left(\pi_D\frac{d D}{dt}+N{\cal H}_\bot\right)^{(1)}=-\frac{6}{N_D}e^{-3D}\sqrt{f}({\rm\bar\Psi}+3{\rm\Phi})
\left(\frac{d D}{dt}\right)^2 Y_{\bf k}^{(s)}-\nonumber\\
&-&\frac{2}{a_0^2}N_De^{-D}\sqrt{f}(3{\cal K}{\rm\bar\Psi}-2k^2{\rm\Phi})Y_{\bf k}^{(s)}+
N_D e^{-3D}\sqrt{f}\rho({\rm\bar\Psi}-3{\rm\Phi})Y_{\bf k}^{(s)}.\nonumber
\end{eqnarray}
Then, taking into account $\sqrt{f}=a_0^3\sqrt{({}^1f)}$ and (\ref{Sigma0}),
after integrating over the space, we obtain the first order correction to the action (\ref{Wh})
\begin{eqnarray}\nonumber
W^{(1)}&=&6 V_0
\int\limits_{t_I}^{t_0}\, dt \frac{e^{-3D}}{N_D}({\rm\bar\Psi}+3{\rm\Phi})\left(\frac{d D}{dt}\right)^2 Y_0^{(s)}+\\
&+&V_0\int\limits_{t_I}^{t_0}\, dt N_D\left[\frac{6 e^{-D}}{a_0^2}{\cal K}{\rm\bar\Psi}-
e^{-3D}\rho({\rm\bar\Psi}-3{\rm\Phi})\right]Y^{(s)}_0.\label{WH1}
\end{eqnarray}
In force of orthogonality of the basis functions (\ref{ort}), there are left only
zero harmonics $Y^{(s)}_0$ in (\ref{WH1})
$$\int\limits_{\Sigma}\, d^3x\sqrt{({}^1 f)}Y_{\bf k}^{(s)}=Y^{(s)}_0.$$
In the expression (\ref{WH1}), there was utilized a condition
$$\int\limits_{\Sigma}\, d^3x\sqrt{({}^1 f)}k^2 Y_{\bf k}^{(s)}=0.$$

Adding (\ref{Whint}) and (\ref{WH1}) $W=W^{(0)}+W^{(1)}$, one yields
\begin{eqnarray}
W=&-&6V_0\int\limits_{t_I}^{t_0} dt\frac{e^{-3D}}{N_D}\left(\frac{d D}{dt}\right)^2
\left[1-({\rm\bar\Psi}+3{\rm\Phi}) Y_0^{(s)}\right]+\nonumber\\
&+&\frac{6V_0}{a_0^2}{\cal K}\int\limits_{t_I}^{t_0} dt{N_D}e^{-D}\left[1+
{\rm\bar\Psi} Y_0^{(s)}\right]-\nonumber\\
&-&V_0\rho\int\limits_{t_I}^{t_0} dt{N_D}e^{-3D}\left[1+({\rm\bar\Psi}-3{\rm\Phi})Y_0^{(s)}\right].\label{WHf}
\end{eqnarray}
Introducing the momentum $p_D$ by the rule
$$
p_D=\frac{\delta W}{\delta\dot D}=-\frac{12V_0}{N_D}e^{-3D}\left(\frac{d D}{dt}\right)
\left[1-({\rm\bar\Psi}+3{\rm\Phi})Y_0^{(s)}\right],$$
we rewrite the action (\ref{WHf}) in the Hamiltonian form
\begin{equation}\label{WHHam}
W=\int\limits_{t_I}^{t_0}\, dt\left[p_D\frac{d D}{dt}+N_D {H}_\bot\right],
\end{equation}
where the Hamiltonian constraint is a following
\begin{eqnarray}\nonumber
&&{H}_\bot=\frac{6V_0}{N_D^2}e^{-3D}\left(\frac{d D}{dt}\right)^2
\left[1-({\rm\bar\Psi}+3{\rm\Phi}) Y_0^{(s)}\right]+\\
&+&\frac{3V_0}{a_0^2}{\cal K} e^{-D}\left[1+{\rm\bar\Psi} Y_0^{(s)}\right]-
V_0\rho e^{-3D}\left[1+({\rm\bar\Psi}-3{\rm\Phi}) Y_0^{(s)}\right].\nonumber
\end{eqnarray}
Thus we obtain the corrected Friedmann equation (\ref{Hubble0})
\begin{equation}\label{Hubblemod}
{H^2\equiv\left(\frac{d D}{dt}\right)^2=\frac{1}{6}N_D^2\rho_{mod},}
\end{equation}
where additions to the density and to the term of curvature are appeal
\begin{equation}\label{rhomod}
{\rho_{mod}=\rho \left[1+2{\rm\bar\Psi}Y_0^{(s)}\right]+
\left[\rho_{curv}-\frac{6{\cal K}}{a^2}(2{\rm\bar\Psi}+3{\rm\Phi}) Y_0^{(s)}\right].}
\end{equation}
To make recalculation in the conformal variables, one should replace the density in
(\ref{rhomod}) according to the conformal transformation:
$$\tilde\rho=\exp{(-4D})\rho.$$
Thus, considering the first perturbation correction of the metric, we obtain the corrected energy of the
Universe and the modified Friedmann equation.
Note, that only zero harmonics of the perturbations considered are included in the (\ref{rhomod}).

$\bullet$
Vector harmonic functions obey to the equation
$$(({}^1\bar{\rm\Delta})+k^2)Y_{{\bf k},i}^{(v)}=0.$$
They are divergent-free
$$({}^1 \bar\nabla_i) Y_{\bf k}^{(v)i}=0.$$
Tensor perturbations are constructed as
$$Y_{{\bf k},ij}^{(v)}\equiv -\frac{1}{2k}\left(({}^1\bar\nabla_i)
Y_{{\bf k},j}^{(v)}+({}^1\bar\nabla_j) Y_{{\bf k},i}^{(v)}\right).$$
Linear perturbations of the metric are taking of the form
\begin{equation}\label{gmnvec}
(g_{\mu\nu})=a^2(t)
\left(
\begin{array}{cc}
    -1&0\\
0&({}^1f_{ij})
\end{array}
\right)+a^2(t)
\left(
\begin{array}{cc}
    0&-B Y_{{\bf k},i}^{(v)}\\
-B Y_{{\bf k},j}^{(v)}&2H_T Y_{{\bf k},{i j}}^{(v)}
\end{array}
\right)
\end{equation}
with some arbitrary functions of time $B(t)$, $H_T(t)$.
The determinant of the spatial metric is
\begin{equation}\label{detvector}
{\det}(\gamma_{ij})=a^6(t).
\end{equation}
The traceless property of $Y_{{\bf k},i}^{(v)i}=0$ under the calculation of (\ref{detvector}) was used.
The inverse metric tensor is
\begin{equation}\label{gmnvecinverse}
(g^{\mu\nu})=\frac{1}{a^2(t)}
\left(
\begin{array}{cc}
    -1&0\\
0&({}^1 f^{ij})
\end{array}
\right)+\frac{1}{a^2(t)}
\left(
\begin{array}{cc}
    0&-B Y_{\bf k}^{(v)i}\\
-B Y_{\bf k}^{(v)j}&-2H_T Y_{\bf k}^{(v)ij}
\end{array}
\right).
\end{equation}

$\bullet$
Tensor harmonic functions obey to the equation on eigenvalues:
$$(({}^1\bar{\rm\Delta})+k^2)Y_{{\bf k},ij}^{(t)}=0.$$
They have properties of traceless
$$Y_{{\bf k},i}^{(t)i}=0,$$
and divergence-free
$$({}^1\bar\nabla_j) Y_{\bf k}^{(t)ij}=0.$$

Perturbations of the metric are presented as
\begin{equation}\label{gmnt}
(g_{\mu\nu})=a^2(t)
\left(
\begin{array}{cc}
    -1&0\\
0&({}^1 f_{ij})
\end{array}
\right)+a^2(t)
\left(
\begin{array}{cc}
    0&0\\
0&2H_T Y_{{\bf k},i j}^{(t)}
\end{array}
\right).
\end{equation}
The determinant of the spatial metric is
\begin{equation}\label{dettensor}
\det (\gamma_{ij})=a^6(t),
\end{equation}
in force of the property $Y_{{\bf k},i}^{(t)i}=0$.
The inverse metric tensor is
\begin{equation}\label{gmnteninverse}
(g^{\mu\nu})=\frac{1}{a^2(t)}
\left(
\begin{array}{cc}
    -1&0\\
0&({}^1f^{ij})
\end{array}
\right)+\frac{1}{a^2(t)}
\left(
\begin{array}{cc}
    0&0\\
0&-2H_T Y_{\bf k}^{(t)ij}
\end{array}
\right).
\end{equation}

We see that in force of (\ref{detvector}), (\ref{dettensor}),
the vector and tensor perturbations in linear approximation do not influence the intrinsic time $D$.

\section{Friedmann equation in Classical\\ cosmology}

A global time exists in homogeneous cosmological models (see, for example, papers \cite{Kasner,Misner}).
The conformal metric $(\tilde\gamma_{ij})$ \cite{PP} for three-dimensional sphere in spherical coordinates
$(\chi, \theta, \varphi)$ is defined via the first quadratic form (\ref{back}).
The intrinsic time $D$ is defined with minus as logarithm of ratio of scales
$$D=-\ln\left(\frac{a(t)}{a_0}\right).$$

In the Standard cosmological model the Friedmann equation is used for fitting SNe Ia data. It
ties the intrinsic intervals time with the coordinate time one
\begin{equation}\label{Friedmannclass}
\left(\frac{dD}{dt}\right)^2\equiv
\left(\frac{\dot{a}}{a}\right)^2=H_0^2\left[\Omega_{\rm M}\left(\frac{a_0}{a}\right)^3+\Omega_{\Lambda}\right].
\end{equation}
Three cosmological parameters favor for modern astronomical observations
$$H_0=h\cdot 10^5 m/s/Mpc,\quad h=0.72\pm 0.08$$
-- Hubble constant,
$$\Omega_\Lambda = 0.72,\quad \Omega_{\rm M}=0.28$$
-- partial densities. Here $\Omega_{\rm M}$ is the baryonic density parameter, $\Omega_\Lambda$ is the density parameter corresponding to $\Lambda$-term, constrained with $\Omega_{\rm M}+\Omega_\Lambda =1.$

The solution of the Friedmann equation (\ref{Friedmannclass}) is presented in analytical form
\begin{equation}\label{classicalsolution}
a(t)=a_0\sqrt[3]{\frac{\Omega_{\rm M}}{\Omega_\Lambda}}
\left[{\rm sinh}\left(\frac{3}{2}\sqrt{\Omega_\Lambda}H_0 t)\right)\right]^{2/3}.
\end{equation}
Here $a(t)$ is a scale of the model, $a_0=1$ is its modern value.
The second derivative of the scale factor is
\be
\ddot{a}=\frac{H_0^2 a_0}{2}\left[2\Omega_\Lambda\left(\frac{a}{a_0}\right)-\Omega_{\rm M}\left(\frac{a_0}{a}\right)^2\right].
\ee
In the modern epoch the Universe expands with acceleration, because $2\Omega_\Lambda >\Omega_{\rm M};$
in the past, its acceleration is negative $\ddot{a}<0.$ This change of sign of the acceleration without clear physical reason puzzles researchers. From the solution (\ref{classicalsolution}), if one puts
$$\frac{a(t)}{a_0}=\frac{1}{1+z},$$
the {\it age -- redshift relation} is followed
\be
H_0t=\frac{2}{3\sqrt{\Omega_\Lambda}}{\rm Arcsinh}
\left(\sqrt{\frac{\Omega_\Lambda}{\Omega_M}}\frac{1}{(1+z)^{3/2}}\right).\label{ageredshiftclassic}
\ee
The age $t_0$ of the modern Universe is able to be obtained by taking $z=0$ in (\ref{ageredshiftclassic})
\begin{equation}
t_0=\frac{2}{3\sqrt{\Omega_\Lambda}}\frac{1}{H_0}{\rm Arcsinh}\sqrt{\frac{\Omega_\Lambda}{\Omega_{\rm M}}}.
\end{equation}

Since for light
$$ds^2=-c^2dt^2+a^2(t)dr^2=0,\quad cdt=-a(t)dr,$$
we have, denoting $x\equiv a/a_0$,
\be\label{int}
-a_0 r=c\int\frac{dt}{x}=c\int\frac{dx}{x}\frac{1}{dx/dt}.
\ee
Rewriting the Friedmann equation (\ref{Friedmannclass}), one obtains a quadrature
\be\label{Friedmannx}
\frac{dx}{dt}=H_0\sqrt{\Omega_{\rm M}/x+\Omega_\Lambda x^2}.
\ee
Substituting the derivative (\ref{Friedmannx}) into (\ref{int}), we get the integral
\be\label{auxiliary}
H_0 r=\frac{c}{\sqrt{\Omega_\Lambda}}\int\limits_{1/(1+z)}^1\frac{dx}{\sqrt{x^4+ 4a_3 x}},
\ee
where we denoted a ratio as
$$4a_3\equiv \frac{\Omega_{\rm M}}{\Omega_\Lambda}.$$
Then we introduce a new variable $y$ by the following substitution
\be\label{xy}
\sqrt{x^4+4a_3 x}\equiv x^2-2y.
\ee
Raising both sides of this equation in square, we get
\be\label{xysquare}
a_3 x=-x^2 y+y^2.
\ee
Differentials of both sides of the equality (\ref{xysquare}) can de expressed in the form:
$$\frac{dx}{x^2-2y}=-\frac{dy}{2xy+a_3}.$$
Utilizing the equality (\ref{xy}), one can rewrite it
\be\label{du}
\frac{dx}{\sqrt{x^4+4a_3x}}=-\frac{dy}{2xy+a_3}.
\ee
Then, we take the expression from the equation (\ref{xysquare})
$$2xy+a_3=\pm\sqrt{4y^3+a_3^2},$$
where a sign plus if
$$0\le x\le\sqrt[3]{\frac{a_3}{2}}=\frac{1}{2}\sqrt[3]{\frac{\Omega_{\rm M}}{\Omega_\Lambda}},$$
and a sign minus if
$$\sqrt[3]{\frac{a_3}{2}}=\frac{1}{2}\sqrt[3]{\frac{\Omega_{\rm M}}{\Omega_\Lambda}}\le x\le 1,$$
and substitute it into the right hand of the differential equation (\ref{du}).
The equation takes the following form
\be
\frac{dx}{\sqrt{x^4+4a_3 x}}=\mp\frac{dy}{\sqrt{4y^3+a_3^2}}\equiv
\mp\frac{dy}{2\sqrt{(y-e_1)(y-e_2)(y-e_3)}},\label{W}
\ee
where
$$y\equiv\frac{1}{2}\left(x^2-\sqrt{x^4+4a_3 x}\right),$$
with three roots:
\bea
e_1&\equiv& \frac{1}{8}\left(\frac{\Omega_{\rm M}}{\Omega_\Lambda}\right)^{2/3}(1+\imath\sqrt{3}),\quad
e_2\equiv -\frac{1}{4}\left(\frac{\Omega_{\rm M}}{\Omega_\Lambda}\right)^{2/3},\nonumber\\
e_3&\equiv& \frac{1}{8}\left(\frac{\Omega_{\rm M}}{\Omega_\Lambda}\right)^{2/3}(1-\imath\sqrt{3}).\label{e1e2e3}
\eea
The integral (\ref{auxiliary}) for the interval $\sqrt[3]{a_3/2}\le x\le 1$, corresponding to
\begin{equation}\label{interval}
0\le z\le 2\sqrt[3]{\frac{\Omega_\Lambda}{\Omega_{\rm M}}}\approx 1.74,
\end{equation}
gives the {\it coordinate distance  - redshift relation} in integral form
\be
H_0 r=
\frac{c}{\sqrt{\Omega_\Lambda}}\int\limits_{[1-\sqrt{1+4a_3(1+z)^3}]/(2(1+z)^2)}^\infty
\frac{dy}{\sqrt{4y^3+a_3^2}}-
\frac{c}{\sqrt{\Omega_\Lambda}}\int\limits_{(1-\sqrt{1+4a_3})/2}^\infty
\frac{dy}{\sqrt{4y^3+a_3^2}}.\label{WP}
\ee
The interval considered in (\ref{interval}) covers the modern cosmological observations one \cite{Riess2004} up to the right latest achieved redshift limit $z\sim 1.7$.

The integrals in (\ref{WP}) are expressed with use of inverse Weierstrass $\wp$-function \cite{Whittaker}
\bea
H_0 r&=&-\frac{c}{\sqrt{\Omega_\Lambda}}\wp^{-1}
\left[\frac{1-\sqrt{1+\Omega_{\rm M}/\Omega_\Lambda(1+z)^3}}{2(1+z)^2}\right]+\nonumber\\
&+&\frac{c}{\sqrt{\Omega_\Lambda}}
\wp^{-1}\left[\frac{(1-\sqrt{1+\Omega_{\rm M}/\Omega_\Lambda})}{2}\right].\label{relationP}
\eea
The invariants of the Weierstrass functions are
$$g_2=0,\qquad g_3=-a_3^2=-\left(\frac{\Omega_{\rm M}}{4\Omega_\Lambda}\right)^2;$$
the discriminant is negative
$$\Delta\equiv g_2^3-27g_3^2<0.$$

Let us rewrite the relation (\ref{relationP}) in implicit form between the variables with use of $\wp$-function
$$\wp u=\frac{1-\sqrt{1+\Omega_{\rm M}/\Omega_\Lambda (1+z)^3}}{2(1+z)^2},$$
where
$$u\equiv \frac{1}{c}\sqrt{\Omega_\Lambda}H_0 r-\wp^{-1}\left(\frac{1-\sqrt{1+\Omega_{\rm M}/\Omega_\Lambda}}{2}\right),$$
The Weierstrass $\wp$-function can be expressed through an elliptic Jacobi cosine function \cite{Whittaker}
\be\label{cnfrac}
\wp u=e_2+H\frac{1+{\rm cn} \left(2\sqrt{H}u\right)}{1-{\rm cn}\left(2\sqrt{H}u\right)},
\ee
where, the roots from (\ref{e1e2e3}) are presented in the form
$$e_1=m+\imath n,\quad e_2=-2m,\quad e_3=m-\imath n,$$
therefore
$$m\equiv \frac{1}{8}\left(\frac{\Omega_{\rm M}}{\Omega_\Lambda}\right)^{2/3},\quad
n\equiv\frac{\sqrt{3}}{8}\left(\frac{\Omega_{\rm M}}{\Omega_\Lambda}\right)^{2/3},$$
and $H$ is calculated according to the rule
$$H\equiv\sqrt{9m^2+n^2}=\frac{\sqrt{3}}{4}\left(\frac{\Omega_{\rm M}}{\Omega_\Lambda}\right)^{2/3}.$$
Then, from (\ref{cnfrac}) we obtain an implicit dependence between the variables, using Jacobi cosine function
\be\label{cnJacobi}
{\rm cn} \left[\sqrt[4]{3}(\Omega_{\rm M}/\Omega_\Lambda)^{1/3}u\right]=\frac{f(z)-1}{f(z)+1},
\ee
where we introduced the function of redshift
\be
f(z)\equiv
\frac{2}{\sqrt{3}}\left(\frac{\Omega_\Lambda}{\Omega_{\rm M}}\right)^{2/3}
\frac{1-\sqrt{1+\Omega_{\rm M}/\Omega_\Lambda(1+z)^3}}{(1+z)^2}+\frac{1}{\sqrt{3}}.\nonumber
\ee
The modulo of the elliptic function (\ref{cnJacobi}) is obtained by the following rule \cite{Whittaker}:
$$k\equiv\sqrt{\frac{1}{2}-\frac{3e_2}{H}}=\sqrt{\frac{1}{2}+\sqrt{3}}.$$

Claudio Ptolemy classified the stars visible to the naked eye into six classes according to their brightness.
The magnitude scale is a logarithmic scale, so that a difference of 5
magnitudes corresponds to a factor of 100 in luminosity. The absolute magnitude $M$ and the apparent magnitude
$m$ of an object are defined as
\be
M\equiv -\frac{5}{2} {\rm lg} \frac{L}{L_0},\qquad
m\equiv -\frac{5}{2} {\rm lg} \frac{l}{l_0},\nonumber
\ee
where $L_0$ and $l_0$ are reference luminosities. In astronomy, the radiated power $L$ of a star or a galaxy,
is called its absolute luminosity. The flux density $l$ is called its apparent luminosity. In Euclidean
geometry these are related as
$$l=\frac{L}{4\pi d^2},$$
where $d$ is our distance to the object. Thus one defines the {\it luminosity distance} $d_L$ of an object as
\begin{equation}\label{dL}
d_L\equiv\sqrt{\frac{L}{4\pi l}}.
\end{equation}
In Friedmann -- Robertson -- Walker cosmology the absolute luminosity
\begin{equation}\label{L}
L=\frac{N_\gamma E_{\rm em}}{t_{\rm em}},
\end{equation}
where $N_\gamma$ is a number of photons emitted, $E_{\rm em}$ is their average energy,
$t_{\rm em}$ is emission time. The apparent luminosity is expressed as
\begin{equation}\label{l}
l=\frac{N_\gamma E_{\rm abs}}{t_{\rm abs}A},
\end{equation}
where $E_{\rm abs}$ is their average energy, and
$$A=4\pi a_0^2 r^2$$
is an area of the sphere around a star. The number of photons is conserved, but their energy is redshifted,
\begin{equation}\label{Eabs}
E_{\rm abs}=\frac{E_{\rm em}}{1+z}.
\end{equation}
The times are connected by the relation
\begin{equation}\label{tabs}
t_{\rm abs}=(1+z)t_{\rm em}.
\end{equation}
Then, with use of (\ref{Eabs}), (\ref{tabs}), the apparent luminosity (\ref{l}) can be presented via the absolute luminosity (\ref{L}) as
$$l=\frac{N_\gamma E_{\rm em}}{t_{\rm em}}
\frac{1}{(1+z)^2}\frac{1}{4\pi a_0^2 r^2}=\frac{1}{(1+z)^2}\frac{L}{4\pi a_0^2 r^2}.$$
From here, the formula for luminosity distance (\ref{dL}) is obtained
\begin{equation}\label{dLs}
d_L (z)_{SC}=(1+z)a_0 r.
\end{equation}

Substituting the formula for coordinate distance (\ref{relationP}) into (\ref{dLs}), we obtain the analytical expression for the luminosity distance
\bea
&&d_L(z)_{SC}=\nonumber\\
&&\frac{c(1+z)}{H_0\sqrt{\Omega_\Lambda}}\!\left(\!
\wp^{-1}\!\left[\!\frac{(1-\sqrt{1+\Omega_{\rm M}/\Omega_\Lambda})}{2}\right]\!\!-\!\!
\wp^{-1}\!\left[\!\frac{1-\sqrt{1+\Omega_{\rm M}/\Omega_\Lambda(1+z)^3}}{2(1+z)^2}\right]\right).\nonumber
\eea

The modern observational cosmology is based on the Hubble diagram.
{\it The effective magnitude -- redshift relation}
\begin{equation}\label{mMcl}
m(z)-M=5{\rm lg} [d_L(z)_{SC}]+{\cal M},
\end{equation}
is used to test cosmological theories ($d_L$ in units of megaparsecs) \cite{Riess2004}.
Here $m(z)$ is an observed magnitude, $M$ is the absolute magnitude,
and ${\cal M}=25$ is a constant.

\section{Friedmann equation in Classical\\ cosmology with conformal units}

The fit of Conformal cosmological model with $\Omega_{\rm rigid} = 0.755,$ $\Omega_{\rm M}=0.245$
is the same quality approximation
as the fit of the Standard cosmological model with $\Omega_\Lambda = 0.72,$ $\Omega_{\rm M}=0.28$,
constrained with $\Omega_{\rm rigid}+\Omega_\Lambda =1$ \cite{PZakh}.
The parameter $\Omega_{\rm rigid}$ corresponds to a rigid state, where the energy density coincides with the pressure $p=\rho$ \cite{Zeld}.
The energy continuity equation follows from the Einstein equations
$$\dot\rho=-3(\rho+p)\frac{\dot{a}}{a}.$$
So, for the equation of state $\rho=p$, one is obtained the dependence $\rho\sim a^{-6}.$
The rigid state of matter can be formed by a free massless scalar field \cite{PZakh}.

Including executing fitting, we write the {\it conformal Friedmann equation} \cite{PP}
with use of significant conformal partial
parameters, discarding all other insignificant contributions
\be\label{Friedmannconf}
\left(\frac{dD}{d\eta}\right)^2\equiv
\left(\frac{{a}'}{a}\right)^2=
\left(\frac{{\cal H}_0}{c}\right)^2\left[\Omega_{\rm rigid}\left(\frac{a_0}{a^4}\right)+
\Omega_{\rm M}\left(\frac{a_0}{a}\right)\right].
\ee
In the right side of (\ref{Friedmannconf}) there are densities $\rho (a)$ with corresponding conformal weights; in the left side a comma denotes a derivative with respect to conformal time. The conformal Friedmann equation
ties intrinsic time interval with conformal time one.
After introducing new dimensionless variable $x\equiv {a}/{a_0},$
the conformal Friedmann equation (\ref{Friedmannconf}) takes a form
\be
\left(\frac{2c}{\sqrt{\Omega_{\rm M}} {\cal H}_0}\right)^2x^2\left(\frac{dx}{d\eta}\right)^2=
4x^3-g_3\equiv 4(x-e_1)(x-e_2)(x-e_3),\label{Weix}
\ee
where one root of the cubic polynomial in the right hand side (\ref{Weix})
is real, other are complex conjugated
\be
e_1\equiv \sqrt[3]{\frac{\Omega_{\rm rigid}}{\Omega_{\rm M}}}\frac{1+\imath\sqrt{3}}{2},\qquad
e_2\equiv -\sqrt[3]{\frac{\Omega_{\rm rigid}}{\Omega_{\rm M}}},\qquad
e_3\equiv \sqrt[3]{\frac{\Omega_{\rm rigid}}{\Omega_{\rm M}}}\frac{1-\imath\sqrt{3}}{2}.\nonumber
\ee
The invariants are the following
$$g_2=0,\qquad g_3= -\frac{4\Omega_{\rm rigid}}{\Omega_{\rm M}}.$$
where ${\cal H}_0$ is the conformal Hubble constant. The conformal Hubble parameter is defined via the Hubble parameter as
${\cal H}\equiv (a/a_0)H$.
The differential equation (\ref{Weix}) describes an effective problem of classical mechanics --
a falling of a particle with mass $8c^2/(\Omega_M{\cal H}_0^2)$ and
zero total energy in a central field with repulsive potential
$$U(x)=\frac{g_3}{x^2}-4x.$$
Starting from an initial point $x=0$ it reaches a point $x=1$ in a finite time $\eta_0$.
We get an integral from the differential equation (\ref{Weix})
\begin{equation}
\int\limits_{1/(1+z)}^1\frac{x dx}{\sqrt{4x^3-g_3}}=-\frac{\sqrt{\Omega_{\rm M}}{\cal H}_0}{2c}\eta.
\end{equation}

Then, we introduce a new variable $u$ by a rule
\begin{equation}\label{xpitau}
x\equiv\wp(u).
\end{equation}
Weierstrass function $\wp (u)$ \cite{Whittaker} satisfies to the differential equation
$$
\left[\frac{d\wp(u)}{du}\right]^2=4\left[\wp (u)-e_1\right]\left[\wp (u)-e_2\right]\left[\wp (u)-e_3\right],
$$
with
$$\wp (\omega_\alpha)=e_\alpha,\qquad \wp'(\omega_\alpha)=0,\qquad \alpha=1,2,3.$$
The discriminant is negative
$$\Delta\equiv g_2^3-27g_3^2<0.$$
The Weierstrass $\zeta$-function satisfies to conditions of quasi-periodicity
$$\zeta (\tau+2\omega)=\zeta (\tau)+2\eta,\qquad
\zeta (\tau+2\omega')=\zeta (\tau)+2\eta',$$
where
$$\eta\equiv\zeta (\omega),\qquad \eta'\equiv \zeta (\omega').$$

The {\it conformal age -- redshift relationship} is obtained in explicit form
\begin{equation}\label{age}
\eta=\frac{2c}{\sqrt{\Omega_{\rm M}}{\cal H}_0}\left(\zeta\left[\wp^{-1}\left(\frac{1}{1+z}\right)\right]-
\zeta\left[\wp^{-1}(1)\right]\right).
\end{equation}
Rewritten in the integral form the Friedmann equation is known in cosmology as the {\it Hubble law}.
The explicit formula for the {\it age of the Universe} can be obtained
\begin{equation}
\eta_0=\frac{2c}{\sqrt{\Omega_{\rm M}}{\cal H}_0}
\left(\zeta \left[\wp^{-1}(0)\right]-\zeta \left[\wp^{-1}(1)\right]\right).
\end{equation}
An interval of coordinate conformal distance is equal to an interval of conformal time $dr=d\eta$, so we can
rewrite (\ref{age}) as {\it conformal distance -- redshift relation}.

A relative changing of wavelength of an emitted photon corresponds to a relative changing of the scale
$$z=\frac{\lambda_0-\lambda}{\lambda}=\frac{a_0-a}{a},$$
where $\lambda$ is a wavelength of an emitted photon, $\lambda_0$ is a wavelength of absorbed photon.
The Weyl treatment \cite{PP} suggests also a possibility to consider
\begin{equation}\label{WeylCC}
1+z=\frac{m_0 a_0}{[a(\eta) m_0]},
\end{equation}
where $m_0$ is an atom original mass.
Masses of elementary particles, according to Conformal cosmology interpretation (\ref{WeylCC}), become running
$$m(\eta)=m_0 a(\eta).$$

The photons emitted by atoms of the distant stars billions of years ago, remember the size of atoms.
The same atoms were determined by their masses in that long time. Astronomers now compare the spectrum of
radiation with the spectrum of the same atoms on Earth, but with increased since that time.
The result is a redshift $z$ of Fraunhofer spectral lines.

In conformal coordinates photons behave exactly as in Minkowski space.
The time intervals $dt= - a dr$ used in Standard cosmology and the time interval used in Conformal
cosmology $d\eta = - dr$ are different.
The conformal luminosity distance $d_L(z)_{CC}$ is related to the standard luminosity one $d_L(z)_{SC}$ as
\cite{PZakh}
$$d_L(z)_{CC}= (1+z)d_L(z)_{SC}=(1+z)^2r (z),$$
where $r(z)$ is a coordinate distance.
For photons $dr/d\eta=-1,$ so we obtain the explicit dependence:
{\it luminosity distance -- redshift relationship}
\be\label{r(z)conformal}
d_L(z)_{CC}=
\frac{2c(1+z)^2}{\sqrt{\Omega_{\rm M}}{\cal H}_0}
\left(\zeta\left[\wp^{-1}\left(\frac{1}{1+z}\right)\right]-
\zeta\left[\wp^{-1}(1)\right]\right).
\ee

{\it The effective magnitude -- redshift relation} in Conformal cosmology has a form
\begin{equation}\label{mMConf}
m(z)-M=5{\rm lg} [d_L(z)_{CC}]+{\cal M}.
\end{equation}

\section{Comparisons of approaches}

The Conformal cosmological model states that conformal quantities are observable magnitudes.
The Pearson $\chi^2$-criterium was applied in \cite{PZakh}
to select from a statistical point of view the best fitting of Type Ia supernovae data
\cite{Riess2004}.
The rigid matter component $\rho_{\rm rigid}$ in the Conformal model substitutes the $\Lambda$-term of the Standard model. It corresponds to a rigid state of matter, when the energy density is equal to its pressure.
The result of the treatment is: the best-fit of the Conformal model is almost the same quality approximation as
the best-fit of the Standard model.

\begin{figure}[tbp]
\begin{center}
\includegraphics[width=3.1in]{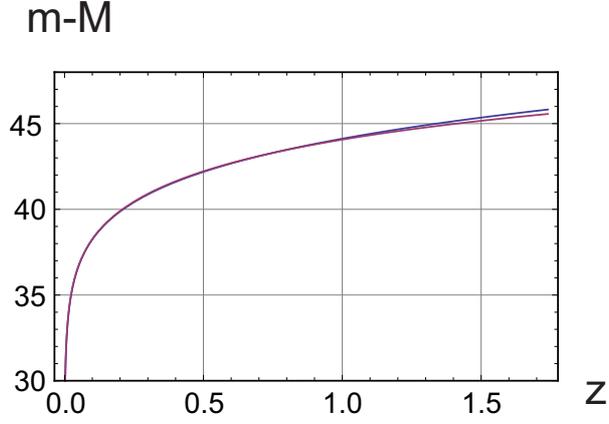}
\caption{\small Curves: the effective magnitude -- redshift relation of the two models.}
\label{Comparison}
\end{center}
\end{figure}
\begin{figure}[tbp]
\begin{center}
\includegraphics[width=3in]{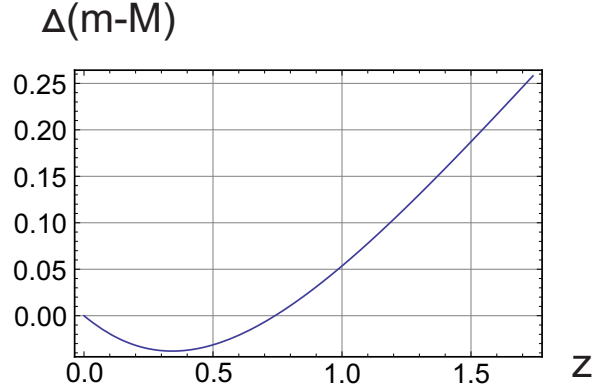}
\caption{\small Difference between curves of the two models: The effective magnitude -- redshift relation.}
\label{Delta}
\end{center}
\end{figure}

Curves of the two models are shown in Fig.\ref{Comparison}.
A fine difference between predictions of the models (\ref{mMConf}) and (\ref{mMcl}):
effective magnitude -- redshift relation
$$\Delta (m(z)-M)=5{\rm lg}[d_L(z)_{CC}]-5{\rm lg}[d_L(z)_{SC}]$$
is depicted in Fig.\ref{Delta}. The differences between the curves are observed in the early and in the past stages of the Universe's evolution.

In Standard cosmology the Hubble, deceleration, jerk parameters are defined as \cite{Riess2004}
\begin{eqnarray}
H(t)&\equiv&+\left(\frac{\dot{a}}{a}\right)=H_0\sqrt{\frac{\Omega_{\rm M}}{a^3}+\Omega_\Lambda},\\
q(t)&\equiv&-\left(\frac{\ddot{a}}{a}\right)\left(\frac{\dot{a}}{a}\right)^{-2}=
\frac{\Omega_{\rm M}/2-\Omega_\Lambda a^3}{\Omega_{\rm M}+\Omega_\Lambda a^3},\\
j(t)&\equiv&+\left(\frac{\dot{\ddot{a}}}{a}\right)\left(\frac{\dot{a}}{a}\right)^{-3}=1.
\end{eqnarray}
As we have seen, the $q$-parameter changes its sign during the Universe's evolution at an inflection point
$$a^{*}=\sqrt[3]{\frac{\Omega_{\rm M}}{2\Omega_\Lambda}},$$
the $j$-parameter is a constant.

We can define analogous parameters in Conformal cosmology also
\begin{eqnarray}
{\cal H}(\eta)&\equiv&+\left(\frac{{a}'}{a}\right),\label{Hparameter}\\
{q}(\eta)&\equiv&-\left(\frac{{a}''}{a}\right)\left(\frac{{a}'}{a}\right)^{-2},\label{qparameter}\\
{j}(\eta)&\equiv&+\left(\frac{{a}'''}{a}\right)\left(\frac{{a}'}{a}\right)^{-3}.\label{jparameter}
\end{eqnarray}
Let us calculate the conformal parameters with use of the conformal Friedmann equation (\ref{Friedmannconf}).
The Hubble parameter
$${\cal H}(\eta)=\frac{{\cal H}_0}{a^2}\sqrt{\Omega_{\rm rigid}+\Omega_{\rm M} a^3}>0;$$
the deceleration parameter
$$q(\eta)=\left(\frac{\Omega_{\rm rigid}-(\Omega_{\rm M}/2)a^3}{\Omega_{\rm rigid}+
\Omega_{\rm M}a^3}\right)>0,$$
so the scale factor grows with deceleration;
the jerk parameter
$$j(\eta)=\frac{3\Omega_{\rm rigid}}{\Omega_{\rm rigid}+\Omega_{\rm M}a^3}>0$$
changes from 3 to $3\Omega_{\rm rigid}$.
The dimensionless parameter $q(\eta)$ and $j(\eta)$ are positive during all evolution. The Universe has not been undergone a jerk.

\section{Conclusions}

In the present paper we have demonstrated that in Geometrodynamics, the many-fingered intrinsic time is a
scalar field. For its construction a background metric was introduced.
To obtain reasonable dynamical characteristics,
in the capacity of background metric one should choose a suitable for corresponding topology a space metric.
For a generic case it is a tangent space, for asymptotically flat problems -- flat one \cite{Gor},
and for cosmological problems -- a compact, or non-compact corresponding manifold.
Hamiltonian approach to obtaining physical observables is not covariant unlike to Lagrangian one.
It seems quite natural: The Lagrangian approach is used to obtain an invariant functional of action,
covariant field equations, suitable for any frame of reference, and the Hamiltonian one --- for getting
physical observables for a given observer in his frame of reference.

The idea of introducing background fields is used traditionally under considering various theoretical problems.
Let us list here some well-known ones. In problems of studying vacuum polarization and quantum particle creation
on curved spaces background fields are necessary for extraction physical quantities \cite{Birrell}.
The procedures of regularization and renormalization of the Casimir vacuum energy are considered in
\cite{GMM,Bordag}. Renormalization involves comparing of some characteristics to obtain as a result of subtraction
the physical observables. For construction of cosmological perturbations theory as a background metric the
Friedmann -- Robertson --Walker one stands \cite{Durrer,Mukhanov}.
The presence of the Minkowski spacetime, as is shown in \cite{Sol}, is necessary to obtain conserved quantities
of gravitational field. The background Minkowski spacetime is presented hidden in asymptotically flat space
problems \cite{ADM}. Topological Casimir energy of quantum fields in closed hyperbolic universes is calculated
in \cite{Muller,Fagundes}.
The problem of obtaining of energy-momentum tensor of the gravitational field in Ricci-flat backgrounds is
discussed in \cite{Grishchuk}.
In Geometrodynamics, the many-fingered intrinsic time is a scalar field.
For obtaining a global time it is not necessary to involve a background metric.

After the York's gauge was implemented,
the deparametrization leads to the global time -- the value of the hypersurface of the Universe.
In application to the problem of the Universe,
the global time is a function of the FRW model scale. It is in agreement with
the stationary Einstein's conception of the Universe \cite{EinsteinCosmology}. The volume of conformal space is constant. Thus we avoid an unpleasant unresolved problem yet of
initial singularity (Big Bang) in the Standard cosmology. The Friedmann equation has a sense of the formula,
connected time intervals (intrinsic, coordinate, conformal) \cite{PP, ZakhP, Pavlovexact}.
If we wish accept the York's extrinsic time,
we get the Friedmann equation as algebraic one. Hence, the connection between temporal intervals
(geometrical coordinate time
$t$ in the pseudo-Riemannian space and intrinsic one $D$ in the WDW superspace) is lost.
Instead of an expansion of the Universe (Standard cosmology) we accept the rate of mass (Conformal cosmology)
\cite{MY}.

In frame of Conformal cosmology it is meaningful to speak of the energy of the Universe that was lost
in Standard cosmology \cite{Wald}.
The some authors (see, for example, \cite{Beluardi, GomesKos, Valentini}) prefer utilizing the York time as a
real time and a volume of the Universe as an operator of evolution.
A linking theory that proves the equivalence of General Relativity and Shape Dynamics was constructed in
\cite{Link}.
In papers \cite{BurlHam, BurlQuantum} there was proposed one to reject of general covariance, so the Hamiltonian constraint is absent. In synchronous system of reference a global time plays a role of global time.
However, the rejection of the Hamiltonian constraint leads to modification of Einstein's gravitation.
The problem of energy in General Relativity naturally tied with the problem of time is concerned as as a basic one during the last century \cite{Hsu}. It was discussed by Hilbert, Noether, Wigner, Dyson, et others.
{\it Most of us have struggled with the problem of how, under these premise, the general theory of relativity can make meaningful statements and predictions at all. Evidently, the usual statements about future positions of particles, as specified by their coordinates, are not meaningful statements in general relativity. This is a point which cannot be emphasized strongly enough and is the basis of a much deeper dilemma than the more technical question of Lorentz invariance of the quantum field equations. It pervades all the general theory, and to some degree we mislead both our students and ourselves when we calculate, for instance, the Mercury perihelion motion without explaining how our coordinates system is fixed in space} \cite{Wigner}. Nowadays, it is quite transparent that general coordinate covariance on which the theory is founded leads to constraints, not to conservation laws. The dynamical problems are solvable in framework of Hamiltonian theory of gravitation.

{\it Nothing is more mysterious and elusive than time. it seems to be the most powerful force in the universe,
carrying us inexorably from birth to death. But what exactly is it? St. Augustine, who died in AD 430, summed
up the problem thus: `If nobody asks me, I know what time is, but if I am asked then I am at a loss what to say'.
All agree that time is associated with change, growth and decay, but is it more than this? Questions abound.
Does time move forward, bringing into being an ever-changing present? Does the past still exist?
Where is the past? Is the future already predetermined, sitting here waiting for us though we know not what
it is?} \cite{Barbour}.
Before $XX$-th century these questions belonged to philosophers. The Einstein's theory
of gravitations allows to stand these questions in physics frame.
The changing volume of the Universe in Standard Cosmology, or changing or masses of elementary particles in Conformal Cosmology is the measure of time, not time is the measure of change.
The so-called expansion of the Universe is able, quite naturally, to be tied with the intrinsic time of the Universe.

As was above demonstrated, Weierstrass and Jacobi functions traditionally used for a long time in classical mechanics and astronomy, are in demand in theoretical cosmology also. The conformal age -- redshift relation, and the effective magnitude -- redshift relations, that are basis formulae for observable cosmology, are expressed explicitly in meromorphic functions. Instead of integral relations, which are used to in cosmology, the derived formulae are expressed through higher transcendental functions, easy to use, because they are built-in analytical software package MATHEMATICA.

The Hubble Space Telescope cosmological supernovae Ia team presented data of high redshifts.
Classical cosmological and Conformal cosmological approaches fit the Hubble diagram with equal accuracy.
According to concepts of Conformal gravitation, conformal quantities of General Relativity are interpreted as physical observables. The conformal cosmological interpretation is preferable because of explaining the resent data without adding the $\Lambda$-term.

It is appropriate to remind the correct statement of the Nobel laureate in Physics Steven Weinberg
\cite{Three} about interpretation of experimental data on redshift. ``{\it I do not want to give the impression
that everyone agrees with this interpretation of the red shift.
We do not actually observe galaxies rushing away from us; all we are sure of is that the lines in their spectra
are shifted to the red, i. e. towards longer wavelengths. There are eminent astronomers who doubt that the
red shifts have anything to do with Doppler shifts or with expansion of the universe}''.

\section*{Acknowledgment}

For fruitful discussions, comments, and criticisms, I would like to thank Prof. V.N. Pervushin.


\end{document}